%% file: main.tex
\documentclass[sigconf,screen]{acmart}

\input{setup}

\title{DepRadar: Agentic Coordination for Context-Aware Defect Impact Analysis in Deep Learning Libraries}

\author{Yi Gao}
\orcid{https://orcid.org/0009-0000-2554-2381}
\affiliation{%
  \institution{The State Key Laboratory of Blockchain and Data Security, Zhejiang University}
  \city{Hangzhou}
  \country{China}
}
\email{gaoyi01@zju.edu.cn}

\author{Xing Hu}
\authornote{Corresponding Author}
\orcid{https://orcid.org/0000-0003-0093-3292}
\affiliation{%
  \institution{The State Key Laboratory of Blockchain and Data Security, Zhejiang University}
  \city{Hangzhou}
  \country{China}
}
\email{xinghu@zju.edu.cn}

\author{Tongtong Xu}
\orcid{https://orcid.org/0000-0002-4323-497X}
\affiliation{%
  \institution{Huawei}
  \city{Hangzhou}
  \country{China}
}
\email{xutongtong9@huawei.com}

\author{Jiali Zhao}
\orcid{https://orcid.org/0009-0007-4389-2925}
\affiliation{%
  \institution{Huawei}
  \city{Hangzhou}
  \country{China}
}
\email{zhaojiali3@huawei.com}

\author{Xiaohu Yang}
\orcid{https://orcid.org/0000-0003-4111-4189}
\affiliation{%
  \institution{The State Key Laboratory of Blockchain and Data Security, Zhejiang University}
  \city{Hangzhou}
  \country{China}
}
\email{yangxh@zju.edu.cn}

\author{Xin Xia}
\orcid{https://orcid.org/0000-0002-6302-3256}
\affiliation{%
  \institution{The State Key Laboratory of Blockchain and Data Security, Zhejiang University}
  \city{Hangzhou}
  \country{China}
}
\email{xin.xia@acm.org}

\copyrightyear{2026}
\acmYear{2026}
\setcopyright{cc}
\setcctype{by}
\acmConference[ICSE '26]{2026 IEEE/ACM 48th International Conference on Software Engineering}{April 12--18, 2026}{Rio de Janeiro, Brazil}
\acmBooktitle{2026 IEEE/ACM 48th International Conference on Software Engineering (ICSE '26), April 12--18, 2026, Rio de Janeiro, Brazil}
\acmPrice{}
\acmDOI{10.1145/3744916.3787763}
\acmISBN{979-8-4007-2025-3/2026/04}

\begin{document}
\input{abstract}

\maketitle

\input{introduction}

\input{motivation}

\input{approach}
\input{evaluation}
\input{discussion}
\input{related_work}
\input{conclusion}

\section*{Acknowledgement}
This research is sponsored by the National Key R\&D Program of China (No. 2024YFB4506400) and CCF-Huawei Populus Grove Fund. We also thank the anonymous reviewers for their insightful comments and suggestions.

\bibliography{main}
\bibliographystyle{ACM-Reference-Format}
\end{document}

%% file: setup.tex
\usepackage[ruled,linesnumbered]{algorithm2e}
\usepackage{multirow}
\usepackage{listings}
\usepackage{booktabs}
\usepackage{multicol}
\usepackage{tabularx}
\usepackage{adjustbox}
\usepackage{subfig}
\usepackage{subcaption}
\usepackage{makecell}
\usepackage{pifont}
\usepackage{enumitem}
\usepackage{calligra}
\usepackage{mdframed}
\usepackage{fontawesome}
\usepackage{amsmath}
\usepackage{amsfonts}
\usepackage{graphicx}
\usepackage{textcomp}
\usepackage{xcolor}
\usepackage{balance}
\usepackage{caption}
\usepackage{wrapfig}
\usepackage{colortbl}
\usepackage{forest}
\usepackage{array}
\usepackage[most]{tcolorbox}

\usepackage{wrapfig}

\usepackage{booktabs}
\usepackage{colortbl}
\usepackage{xcolor}

\usepackage{forest}
\usepackage[most]{tcolorbox}

\def\BibTeX{{\rm B\kern-.05em{\sc i\kern-.025em b}\kern-.08em
    T\kern-.1667em\lower.7ex\hbox{E}\kern-.125emX}}

\lstdefinestyle{codeStyle}{
    basicstyle=\small\ttfamily,
    commentstyle=\color{green},
    keywordstyle=\color{blue},
    stringstyle=\color{blue},
    showstringspaces=false,
    breaklines=true,
    frame=lines,
    backgroundcolor=\color{white},
    captionpos=b,
    aboveskip=10pt,
    belowskip=10pt
}
\definecolor{lightgreen}{rgb}{0.9,1,0.9}
\definecolor{grayheader}{gray}{0.85}
\definecolor{codebackground}{RGB}{240,240,240} %

\newcommand{\appname}{{\sc DepRadar}\xspace}

%% file: abstract.tex
\begin{abstract}
Deep learning (DL) libraries like Transformers and Megatron are now widely adopted in modern AI programs. 
However, when these libraries introduce defects—ranging from silent computation errors to subtle performance regressions—it is often challenging for downstream users to assess whether their own programs are affected. 
Such impact analysis requires not only understanding the defect semantics but also checking whether the client code satisfies complex triggering conditions involving configuration flags, runtime environments, and indirect API usage.
We present \appname, an agent coordination framework for fine-grained defect and impact analysis in DL library updates.
\appname coordinates four specialized agents across three steps:
(1) the \textit{PR Miner} and \textit{Code Diff Analyzer} extract structured defect semantics from commits or pull requests,
(2) the \textit{Orchestrator Agent} synthesizes these signals into a unified defect pattern with trigger conditions, and
(3) the \textit{Impact Analyzer} checks downstream programs to determine whether the defect can be triggered.
To improve accuracy and explainability, \appname integrates static analysis with DL-specific domain rules for defect reasoning and client-side tracing.
We evaluate \appname on 157 PRs and 70 commits across two representative DL libraries. 
It achieves 90\% precision in defect identification and generates high-quality structured fields (average field score 1.6/2). 
On 122 client programs, \appname identifies affected cases with 90\% recall and 80\% precision, substantially outperforming other baselines. 
\end{abstract}

\ccsdesc[500]{Software and its engineering~Software libraries and repositories}

\keywords{Multi-Agent Framework, Pull Request, Defect Impact Analysis}

%% file: introduction.tex
\section{Introduction}
\label{introduction}

Modern deep learning (DL) programs increasingly depend on libraries such as Transformers~\cite{transformers} and Megatron~\cite{megatron}, which streamline model development, training, and deployment~\cite{zeng2024survey,alfadel2023empirical,wu2023understanding,lyu2023chronos,kula2018developers,samarinunderstanding,chen2023toward,pepe2024taxonomy,li2024knowbug,jia2021symptoms,otoom2019automated}. 
These libraries encapsulate critical functionalities such as tokenization, optimizer scheduling, distributed communication, and memory management, enabling developers to build large-scale models efficiently.

However, as these libraries grow in complexity and scale, they unavoidably introduce software defects. 
Bugs in DL libraries may not always manifest as program crashes; instead, they often degrade performance silently, introduce numerical inconsistencies, or corrupt training dynamics under certain conditions~\cite{chen2023toward,yu2025towards}. 
For example, a subtle error in tensor parallelism coordination or configuration parsing may lead to incorrect gradient updates or inefficient resource usage, ultimately impairing model convergence.
In practice, such defects are often discussed in GitHub pull requests (PRs)~\cite{pr} or commits~\cite{commit}, where developers propose fixes based on user feedback or internal audits~\cite{xiao2024generative,herbold2022fine}.
Thus, we can analyze PRs and commits to assess whether a defect impacts downstream client programs.
However, this is highly non-trivial in practice: PRs may reference internal APIs, abstract trigger conditions, or rely on complex runtime configurations. 
Meanwhile, downstream users may not closely monitor upstream libraries and are frequently unaware of whether such defects could impact their own programs.

Although keeping dependencies up-to-date is a common practice for maintaining software reliability, frequent dependency upgrades are impractical in DL environments, as each update may trigger costly retraining and break compatibility~\cite{2024Understanding,2023Toward,2024Bug,10.5555/3767901.3767919}.
Moreover, many defects remain silent across releases or are fixed with minimal documentation.
Release notes are typically coarse (e.g., fix stability) and omit critical details such as affected modules, configurations, or triggering conditions.
For instance, commit \texttt{1674ce3}~\cite{1674ce3} in \textit{Megatron} silently skewed multi-user training for months before detection, illustrating how such fixes can propagate unnoticed.

Recent efforts have leveraged large language models (LLMs) to assist with code summarization and PR triage~\cite{liu2019automatic}. 
However, these approaches typically treat commits or patches in isolation, without modeling the downstream impact on external client programs.
Determining whether a DL library defect affects a specific client program involves several unique challenges:

\noindent\textbf{Challenge 1: Noisy and informal defect representations}.
Defect-related PRs or commits often contain unstructured descriptions, partial fixes, or discussion noise. 
The root cause and triggering conditions are rarely formalized, making the automated extraction of defect patterns difficult.

\noindent\textbf{Challenge 2: Deep semantic gap between library changes and user-visible behavior}.
Many patches fix low-level internal logic (e.g., CUDA graphs, communication buffers), while user programs interact only with high-level APIs. 
Mapping these low-level changes to user-facing risk factors (e.g., configuration parameters, method calls) requires semantic reasoning.

\noindent\textbf{Challenge 3: Semantic alignment between defect conditions and client usage}. 
Determining whether a client program satisfies the defect's triggering conditions requires understanding not only the patched library logic, but also the usage context in the user code—including configurations and function calls. 

To address these challenges, we propose \appname, an agent coordination framework for automated defect and impact analysis in deep learning libraries. 
\appname assists developers in determining whether a defect in an upstream DL library (e.g., Transformers) has concrete effects on their downstream programs. 
Unlike prior approaches that rely on static dependency resolution or flat summarization, \appname leverages a modular agent architecture to reason across both the library side and the client side, enabling context-aware, condition-sensitive impact assessment.
By integrating domain-specific coordination rules, agent self-adaptation, and AST-based structural validation, \appname turns generic multi-agent reasoning into a practical and verifiable workflow for DL defect analysis.
\appname consists of four agents: (1) \textbf{Miner Agent} identifies whether a commit or pull request addresses a real defect and extracts structured defect metadata, including bug background, trigger conditions, and affected components.
(2) \textbf{Code Diff Agent} analyzes the patch semantics, locating modified methods and summarizing the underlying fix logic.
(3) \textbf{Orchestrator Agent} synthesizes the results from upstream agents into a defect pattern, capturing risk-relevant method calls and parameter combinations from the perspective of external users.
(4) \textbf{Impact Analyzer Agent} performs contextual analysis on the client program code to determine whether it satisfies the triggering conditions, guided by static analysis and domain-specific validation rules.

To mitigate the long-context limitations of LLMs, \appname incorporates a progressive context augmentation mechanism that dynamically expands or compresses the input prompt to ensure information completeness without exceeding token limits.
Furthermore, to mitigate hallucinations in LLM-generated outputs, we design a post-processing validation module that uses AST-based static analysis to verify the existence of referenced parameters and methods within the actual client program. 

We evaluate \textbf{\appname} on 157 PRs and 70 commits collected from two widely used DL libraries: \texttt{Transformers} and \texttt{Megatron}.
Our study focuses on two core tasks: (1) whether \appname can accurately generate structured \textit{defect patterns}, and (2) whether it can determine if a downstream client program is semantically affected by a given defect.
On the PR dataset, \appname achieves a defect classification F1-score of 95\%, and over 70\% of extracted defect pattern fields—such as bug background and trigger conditions—are rated fully accurate by domain experts.
In downstream impact analysis across 122 real-world client programs, \appname achieves an F1-score of 85\%, substantially outperforming three strong baselines.
We further evaluate \appname on internal commits of \texttt{Megatron}, where it successfully identifies 12 real-world client programs affected by silent defects, as confirmed by developers.

In summary, the main contributions of this paper include:
\begin{itemize}[leftmargin=*]
\item We present \textbf{\appname}, a novel agent coordination framework for fine-grained defect and impact analysis in deep learning libraries. 
\item We implement a practical system that integrates LLM-based agents with static analysis and domain-specific inference rules. 
The replication package is available at~\cite{depradar}.
\item We conduct an extensive evaluation of 227 library updates (157 PRs and 70 commits) from \texttt{Transformers} and \texttt{Megatron}. 
Results show that \appname outperforms three strong baselines in both defect modeling (F1 = 95\%) and client impact analysis (F1 = 85\%), and identifies real affected clients.
\end{itemize}

%% file: motivation.tex
\section{Motivation Example}
\label{sec:motivation}

\begin{figure}
\centering
\includegraphics[width=\linewidth]{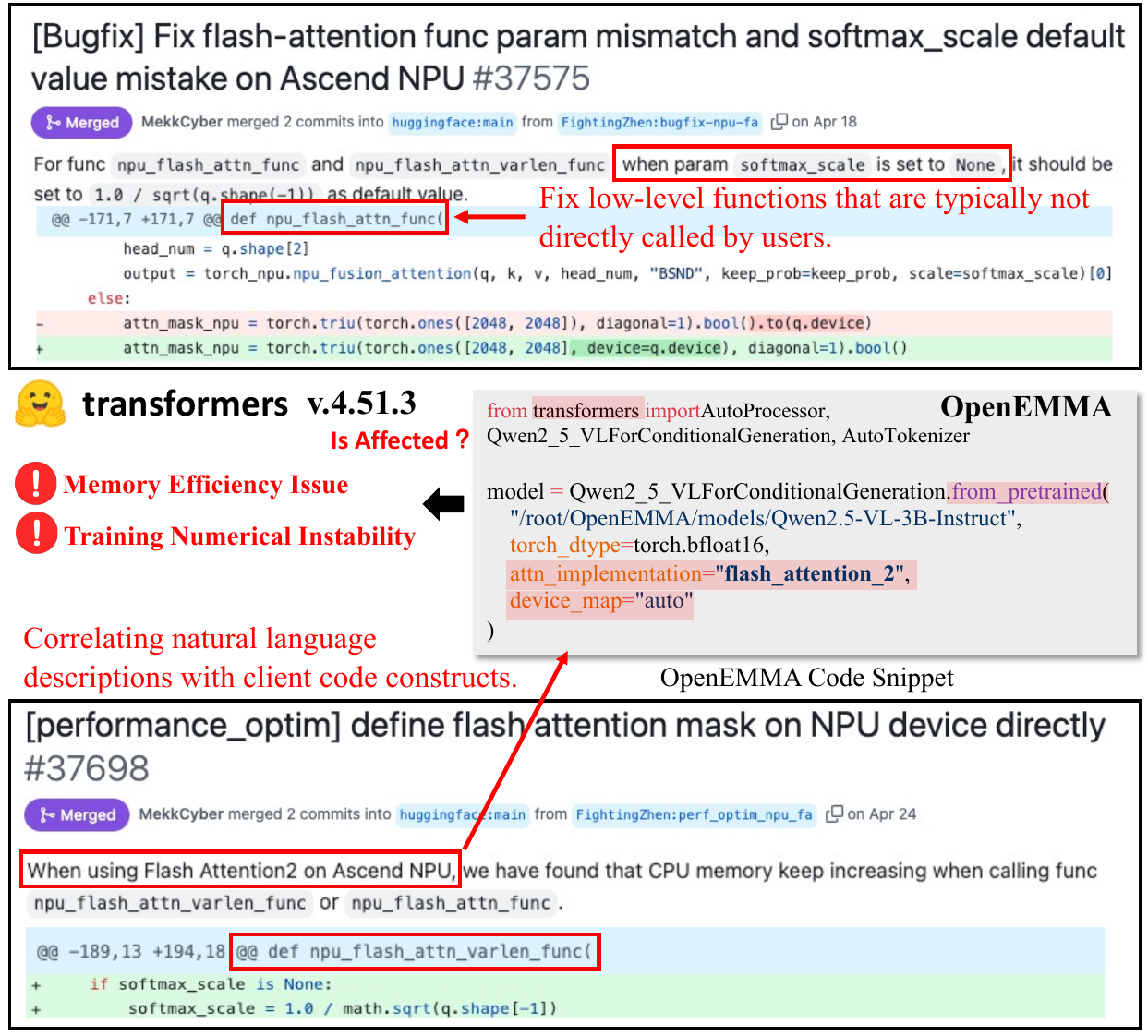}
\caption{Motivation Example.}
\label{fig:moti}
\vspace{-0.5cm}
\end{figure}

\begin{figure*}
\centering
\includegraphics[width=0.9\linewidth]{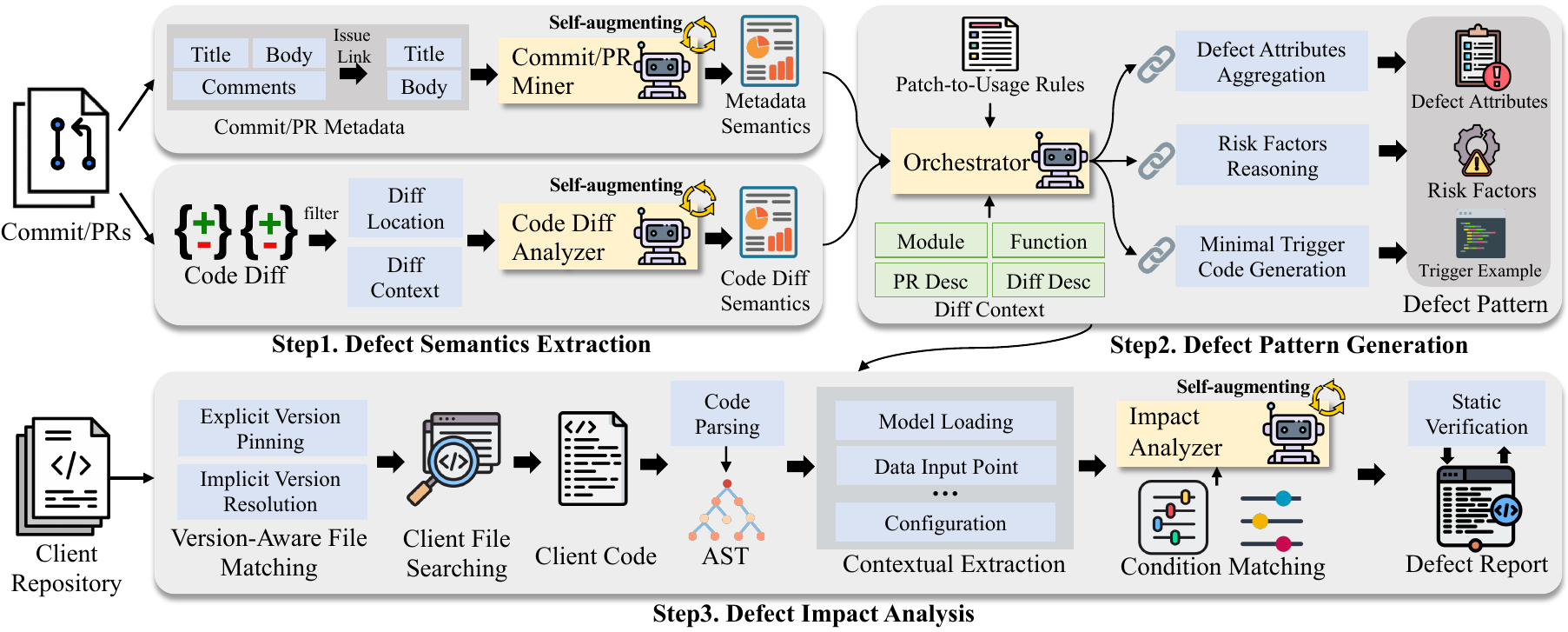}
\caption{Overview of our approach.}
\label{fig:approach}
\vspace{-0.1cm}
\end{figure*}

Given the high cost of silent correctness or performance regressions during model training, automatically identifying library defects and assessing their potential impact on downstream client programs has become an urgent and practical need.
Real-world deep learning libraries like Transformers frequently fix silent defects—bugs that do not crash programs but silently degrade numerical accuracy or runtime efficiency under specific conditions.
However, client developers often lack the visibility and expertise to determine whether such internal fixes are relevant to their own configurations.

Figure~\ref{fig:moti} shows this challenge through two representative PRs from Transformers, both affecting Flash Attention on Ascend NPUs: 
\textbf{PR \#37575} fixes a subtle mistake in the default value of the \texttt{softmax \_scale} parameter. 
If a user enables Flash Attention v2 (\texttt{attn\_imp lementation="flash\_attention\_2"}) without explicitly setting \texttt{softmax\_scale}, 
the resulting attention output may become numerically unstable, harming model quality. 
\textbf{PR \#37698} resolves a memory inefficiency where the attention mask is first initialized on the CPU and then moved to the NPU. 
This silent performance defect causes unnecessary CPU memory allocation and long-term accumulation in certain deployment settings. 

The middle part of Figure~\ref{fig:moti} shows a simplified snippet from a downstream client project (\textit{OpenEMMA, an open-source autonomous driving framework}) that uses Transformers with Flash Attention v2 and NPU hardware—satisfying the trigger conditions of both PRs.
However, neither bug manifests through obvious exceptions or logs, and the developers were unaware of their presence before manual inspection.
Such usage patterns, which involve using pretrained configurations and enabling hardware-optimized features, are commonly seen in downstream DL programs and can make silent defects difficult to detect.

This example highlights a fundamental gap in current DL library usage:
developers cannot feasibly monitor every upstream change or PR, nor can they easily determine if a specific patch is relevant to their setup. 
Furthermore, existing tools either focus on syntax-level patch analysis or general-purpose summarization, lacking the semantic reasoning and domain knowledge needed to bridge the gap between internal bug fixes and client-side manifestations.
These observations motivate our work. 
We propose a novel approach named \appname, analyzes library-level bug-fix PRs and automatically traces their potential runtime impact on client programs.
By combining defect pattern modeling, trigger condition extraction, and client-side reasoning, our framework aims to assist users in identifying whether they are silently affected.


%% file: approach.tex
\section{Approach}
\label{sec:approach}

We propose \textbf{\appname}, an agent coordination framework for automated defect impact analysis in DL libraries.
As shown in Figure~\ref{fig:approach}, \appname consists of three steps: (1) extracting structured defect semantics from PRs/commits via dual agents, (2) synthesizing complete defect patterns, and (3) verifying whether clients satisfy the triggering conditions.
The framework is built on AutoGen~\cite{autogen} to enable coordinated, multi-turn reasoning across agents.

\subsection{Defect Semantics Extraction}
This step involves two specialized agents that collaboratively extract defect semantics from different perspectives: the \textit{Commit/PR Miner Agent} processes natural language descriptions (e.g., commit messages, PR discussions), while the \textit{Code Diff Analyzer Agent} focuses on structural code changes.
To enable multi-agent coordination, we introduce an \textit{Orchestrator Agent} that validates intermediate outputs and synthesizes final defect patterns by combining upstream results.
In this step, the \textbf{Orchestrator Agent} monitors the completeness of extracted semantics.

\begin{figure}
\centering
\includegraphics[width=\linewidth]{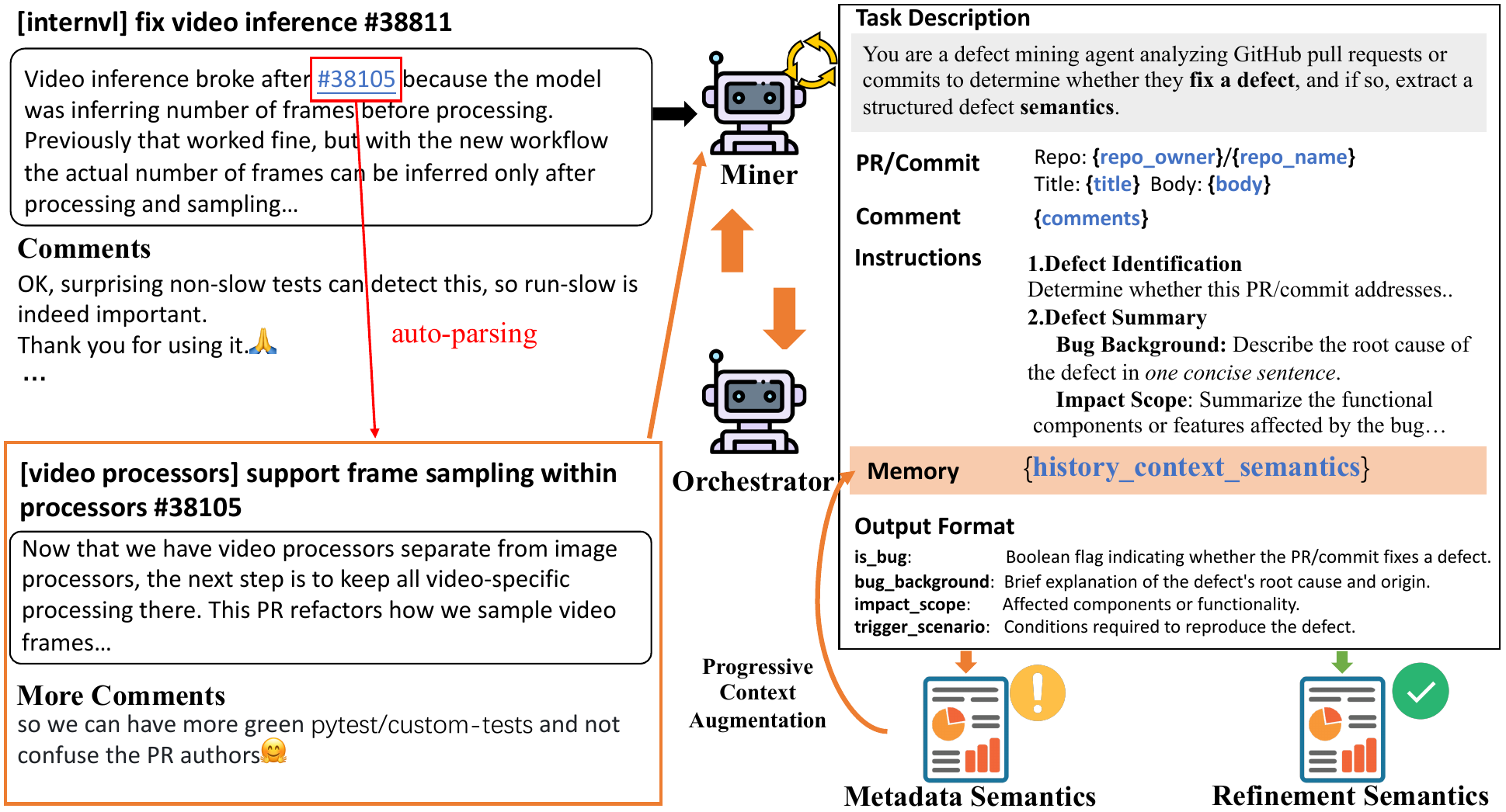}
\caption{Prompt design for the Miner Agent: mining defect patterns from unstructured PR metadata for PR~\#38811 and associated PR~\#38105.}
\label{fig:miner_agent}
\vspace{-0.65cm}
\end{figure}

\subsubsection{Commit/PR Miner Agent}
The Miner Agent is designed to extract structured defect representations from noisy and loosely formatted PR or commit metadata.
Given a PR, \appname processes the title, body, associated issue references, and developer comments to generate a structured defect semantic representation containing three fields: \texttt{bug\_background}, \texttt{impact\_scope}, and \texttt{trigger\_conditions}.

Considering the substantial number of PRs—most of which involve routine updates or feature additions rather than defect fixes, comprehensive analysis of all PRs proves inefficient. 
To address this, we implement a two-step filtering mechanism.
The Miner Agent integrates the GitHub API as a tool to retrieve full metadata for each PR.
It then applies keyword-based heuristics to retain candidates containing defect-related cues (e.g., \textit{fix}, \textit{bug}, \textit{error}) while excluding clearly benign updates (e.g., \textit{doc}, \textit{refactor}, \textit{feat}).
To ensure reliability, we manually inspected all excluded PRs in this period and did not find missed defect-fixing PRs.
The full keyword list and filtering script are released in the reproduction package for transparency~\cite{depradar}.

\noindent \textbf{Progressive Context Augmentation.}
To handle noisy, lengthy, or fragmented PR metadata, \appname adopts a progressive context augmentation strategy coordinated by the Orchestrator Agent.
As illustrated in Figure~\ref{fig:miner_agent}, the Miner Agent processes PR content in fixed-size chunks (default $k=3$ items per round) and generates intermediate summaries capturing partial semantics such as bug background or triggering conditions.
After each round, the Orchestrator Agent validates the content completeness of the extracted fields (e.g., \texttt{trigger\_conditions}, \texttt{bug\_background}).
If the output is insufficient, the Orchestrator Agent  will provide feedback to the Miner Agent. 
The previous semantic representation is preserved as historical context and augmented with the next chunk of PR content.
This iterative refinement continues until all required fields are confidently populated or the full PR metadata is exhausted.
This mechanism ensures that the agent remains within prompt length limits while gradually expanding coverage, enabling robust reasoning over scattered or delayed defect signals.

For example, in PR~\#38811 as shown in Figure~\ref{fig:miner_agent}, although the title reported a \textit{video inference failure} without disclosing root causes or triggering conditions, the Miner Agent coordinated by the Orchestrator—automatically followed linked PR~\#38105 and developer comments to retrieve additional context. 
Through progressive context augmentation, \appname accurately reconstructed the defect pattern, identifying the underlying cause as a frame sampling misalignment in multi-frame video inference pipelines and extracting the specific conditions under which the bug is triggered.

\subsubsection{Code Diff Analyzer Agent}
While PR metadata offers surface-level explanations, the actual bug fix logic is encoded within the code changes.
The Code Diff Agent analyzes these changes to identify repair intent and recover trigger conditions from the changes.
Given a modified patch, we first apply a denoising phase using AST-level parsing to eliminate non-functional edits such as comment updates, formatting, or import reordering.
We retain only structural code changes (e.g., control-flow updates, function/method additions, and validation logic) for downstream analysis.

\begin{figure}
\centering
\includegraphics[width=\linewidth]{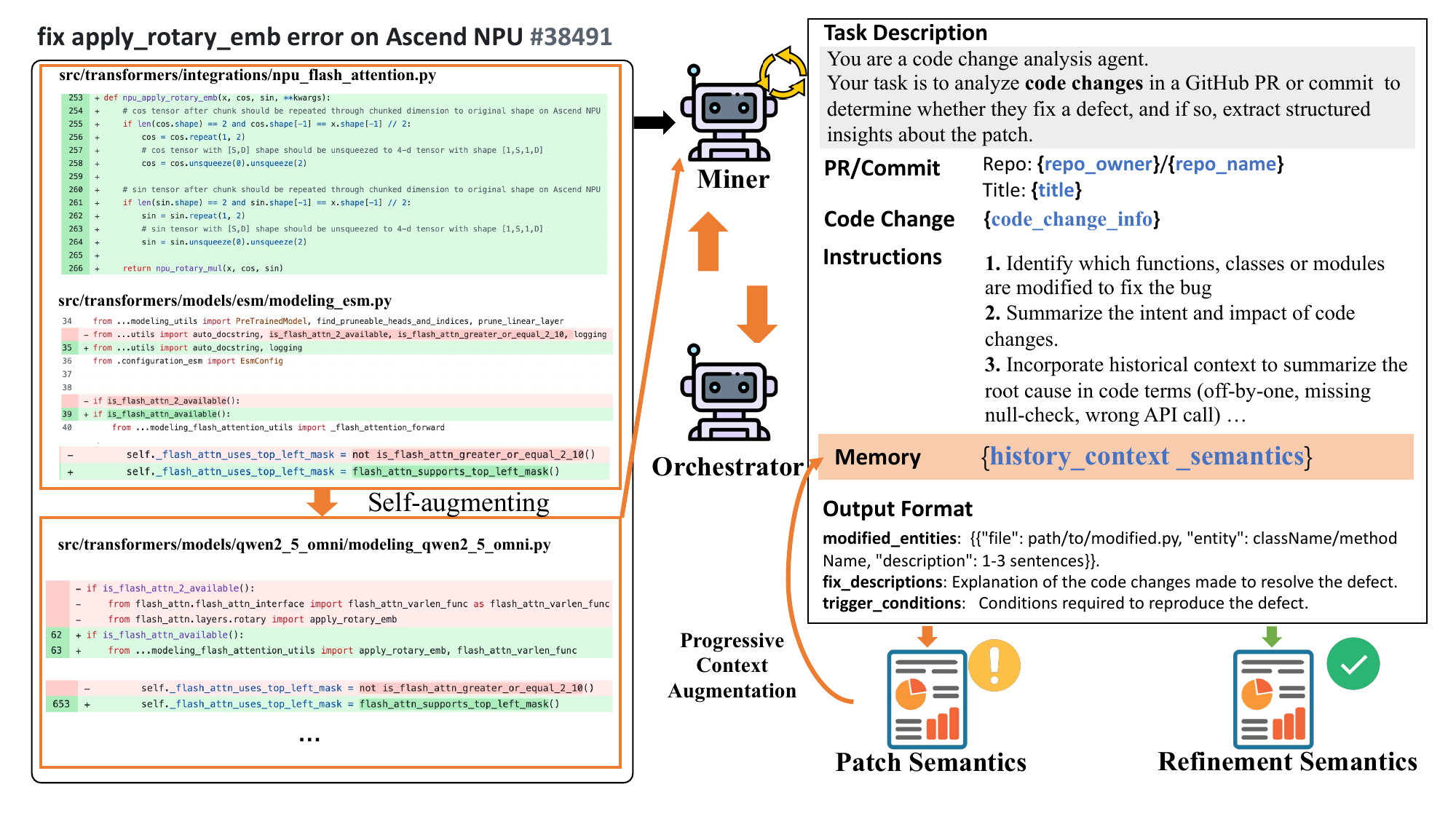}
\caption{Prompt design for the Code Diff Agent: summarizing patch semantics for PR~\#38491.}
\label{fig:codediff_agent}
\vspace{-0.4cm}
\end{figure}

As shown in Figure~\ref{fig:codediff_agent}, the CodeDiff Agent analyzes the code-level patch to extract structured defect semantics from the perspective of program edits. 
Specifically, it captures both the surface fix operations and their underlying rationale, producing a dual-perspective output that complements textual representations from the Miner Agent. 
The extracted semantics consist of three key components: (1) modified code entities annotated with semantic change types (e.g., \textit{added null check}, \textit{modified tensor shape validation}), (2) a synthesized explanation of the root cause underlying the defect (e.g., \textit{missing boundary check}), and (3) a set of key program elements—such as parameters and method names—that are relevant to the defect’s trigger conditions. 
These elements are subsequently used as search keys in the downstream client-side code matching.

\noindent \textbf{Progressive Context Augmentation.}
To process large or multi-file patches where code diffs exceed the input limit of a single LLM prompt, \appname adopts a progressive augmentation strategy similar to that of the Miner Agent. 
As shown in Figure~\ref{fig:codediff_agent}, the CodeDiff Agent segments the patch into logical units (e.g., grouped by file) and analyzes $k=3$ edits per round. 
At each step, the agent generates intermediate summaries that describe fix actions (e.g., API adjustments, parameter constraints) and defect-related semantics.

The Orchestrator Agent then validates whether key semantic fields—such as \texttt{fix\_description}, \texttt{trigger\_conditions} are sufficiently captured. 
If not, the next round’s input includes both new code chunks and the prior semantics. 
This iterative feedback loop continues until either all semantic slots are confidently filled or the full patch is exhausted.
This feedback loop continues until all required semantics are covered or the patch is fully processed.


\begin{table}
\centering
\caption{Domain-Specific Rules for Defect Pattern Inference.}
\label{tab:orchestrator-rules}
\begin{tabular}{p{1.5cm}|p{6cm}}
\toprule
\textbf{Rule Type} & \textbf{Description and Example} \\
\midrule
\textbf{Entity Lifting} & Promote internal symbols (e.g., \texttt{\_flash \_kernel}) to nearest user-facing API (e.g., \texttt{generate()}) based on call graph or naming patterns. \\
\hline
\textbf{Parameter Exposure} & Include only parameters visible in public constructors or config (e.g., \texttt{attention\_impl}). \\
\hline
\textbf{Trigger Synthesis} & Combine condition fragments (e.g., hardware flags, dtype checks) into a minimal executable client snippet. Example: map \texttt{device\_mesh != None} to \texttt{device\_map="auto"}. \\
\hline
\textbf{Semantic Merge} & When multiple triggers appear (e.g., backend flag + precision), enforce conjunction to avoid partial matches. \\
\bottomrule
\end{tabular}
\vspace{-0.5cm}
\end{table}

\subsection{Defect Pattern Generation}
At this step, the Orchestrator combines outputs from prior agents and applies domain rules to generate a structured defect pattern for impact analysis.
Prior work on bug summarization often stops at generating human-readable descriptions or identifying changed APIs~\cite{liu2019automatic,xiao2024generative}.
However, this is insufficient for determining downstream impact.
Many bugs involve internal implementation details or latent behaviors (e.g., device-specific logic) whose consequences are only visible when mapped to concrete usage patterns in clients.

The output follows a structured schema that captures bug-fix semantics from a user-facing perspective.
It contains:(1) Defect Attributes: Whether the patch fixes a bug, its background, and impact.
(2) Risk Factors: Externally visible method names and parameters that can trigger the bug.
(3) Minimal Trigger Example: A synthesized code snippet illustrating the minimal condition required to activate the bug using client-facing APIs.
The Orchestrator Agent integrates upstream outputs and applies rule-based reasoning to transform internal fixes into client-relevant risk factors.

To bridge the semantic gap between low-level fixes in patches and user-facing patterns, we design a domain-informed inference layer guided by a configurable set of rules.
Defect fixes in DL libraries often modify internal methods that users rarely invoke directly; therefore, mapping internal changes to the minimal user-visible trigger scenario is essential for impact analysis.
Our rules steer the Agent to reason beyond raw diffs, identify exposed parameters or configuration keys, and synthesize the smallest usage example that can reproduce the defect in a client context.
These rules are implemented as declarative templates stored in an external configuration file, making them easy to extend for additional libraries or edge cases.
Table~\ref{tab:orchestrator-rules} illustrates representative rule types and their objectives; the full specification and domain-specific variations (e.g., for distributed training flags) are included in our artifact~\cite{depradar}.

\begin{figure}
\centering
\includegraphics[width=0.97\linewidth]{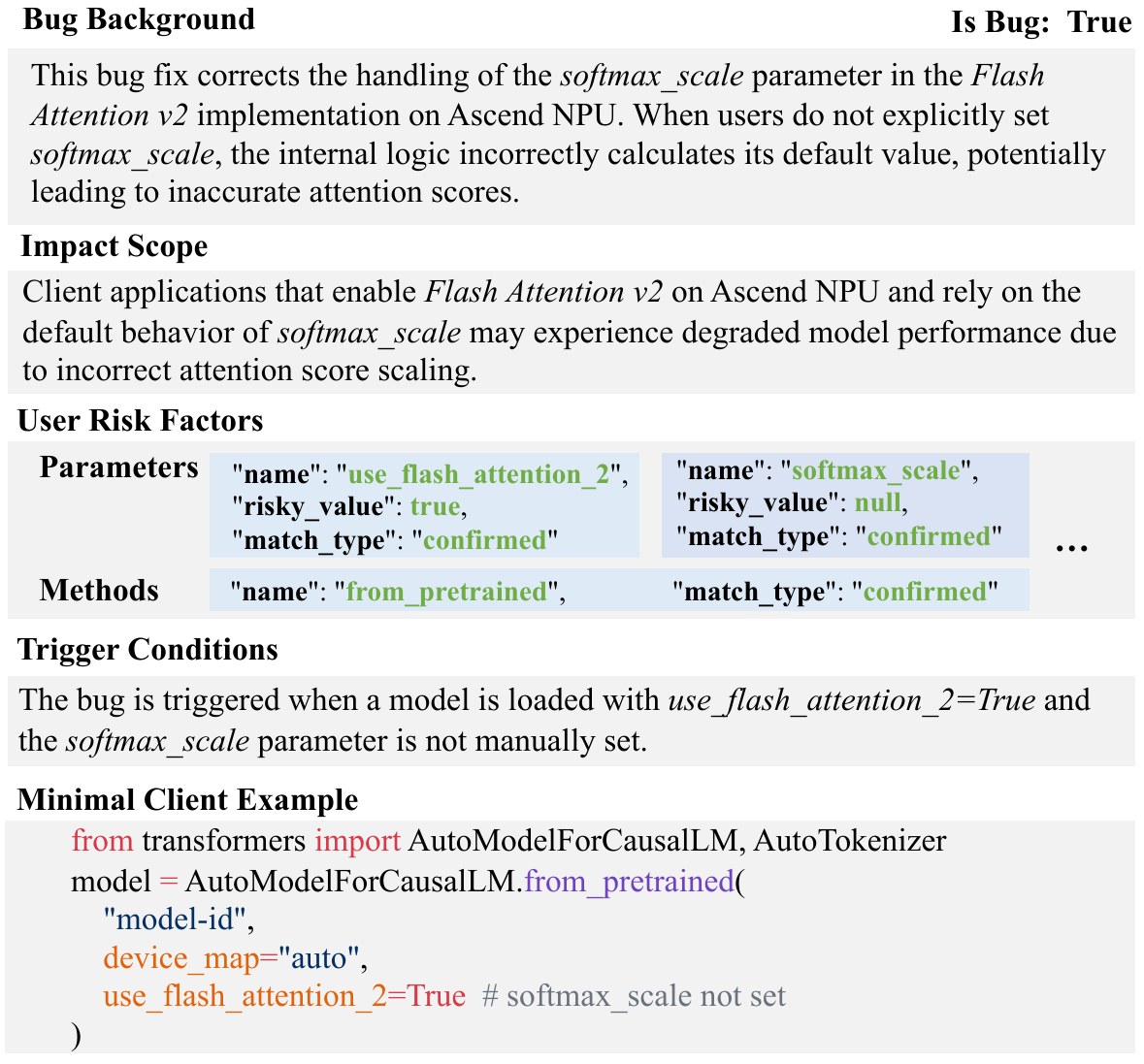}
\caption{Defect Pattern example synthesized for PR~\#37575.}
\label{fig:defect-pattern-example}
\vspace{-0.4cm}
\end{figure}

Figure~\ref{fig:defect-pattern-example} shows the defect pattern generated for PR~\#37575.
This patch fixes an internal bug in Ascend NPU’s Flash Attention implementation, where setting \texttt{softmax\_scale=None} in functions such as \texttt{npu\_flash\_attn\_func} and \texttt{npu\_flash\_attn\_varlen\_ func} leads to incorrect attention scaling.
Although these changes modify low-level device-specific logic, the Orchestrator Agent infers that users who enable Flash Attention v2 without explicitly setting \texttt{softmax\_scale} are silently affected.
Moreover, the generated \texttt{minimal\_client\_example} captures a representative code snippet that satisfies the trigger conditions.
In the subsequent analysis step, this example serves as a search anchor: static analyzers extract parameter names and API calls from the snippet to match against client code contexts.
Overall, the structured defect pattern enables downstream tasks including parameter matching, condition validation, and precise retrieval of affected user code segments.

\begin{figure}
\centering
\includegraphics[width=\linewidth]{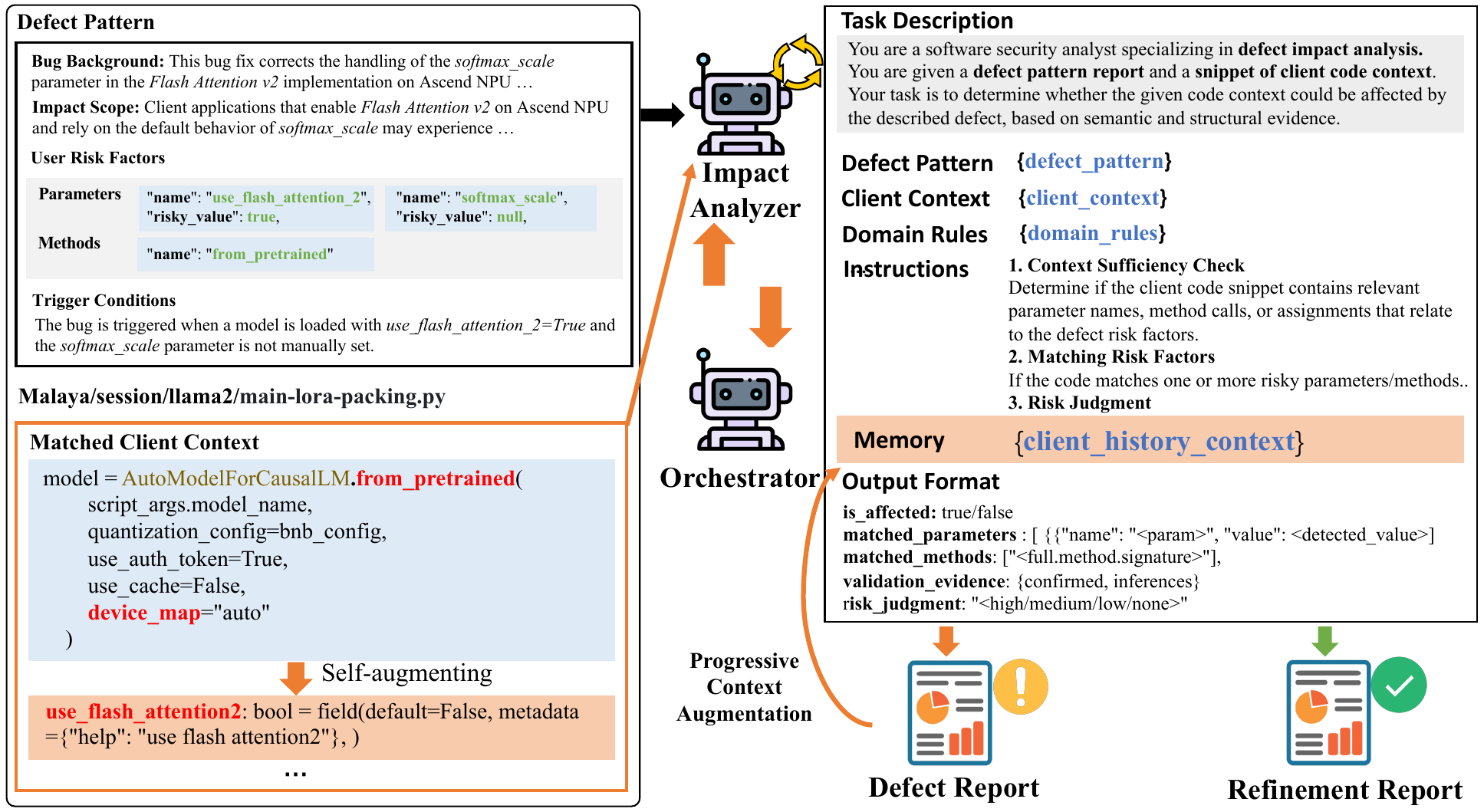}
\caption{Prompt design for the Impact Analyzer Agent: verifying trigger condition satisfaction from matched client code.}
\label{fig:analysis_agent}
\vspace{-0.4cm}
\end{figure}

\subsection{Client-Side Defect Impact Analysis}
In the final step, \appname employs an Impact Analyzer Agent to perform semantic alignment and context-aware reasoning, determining whether the client is affected by the defect.
\begin{figure*}
\centering
\includegraphics[width=0.85\linewidth]{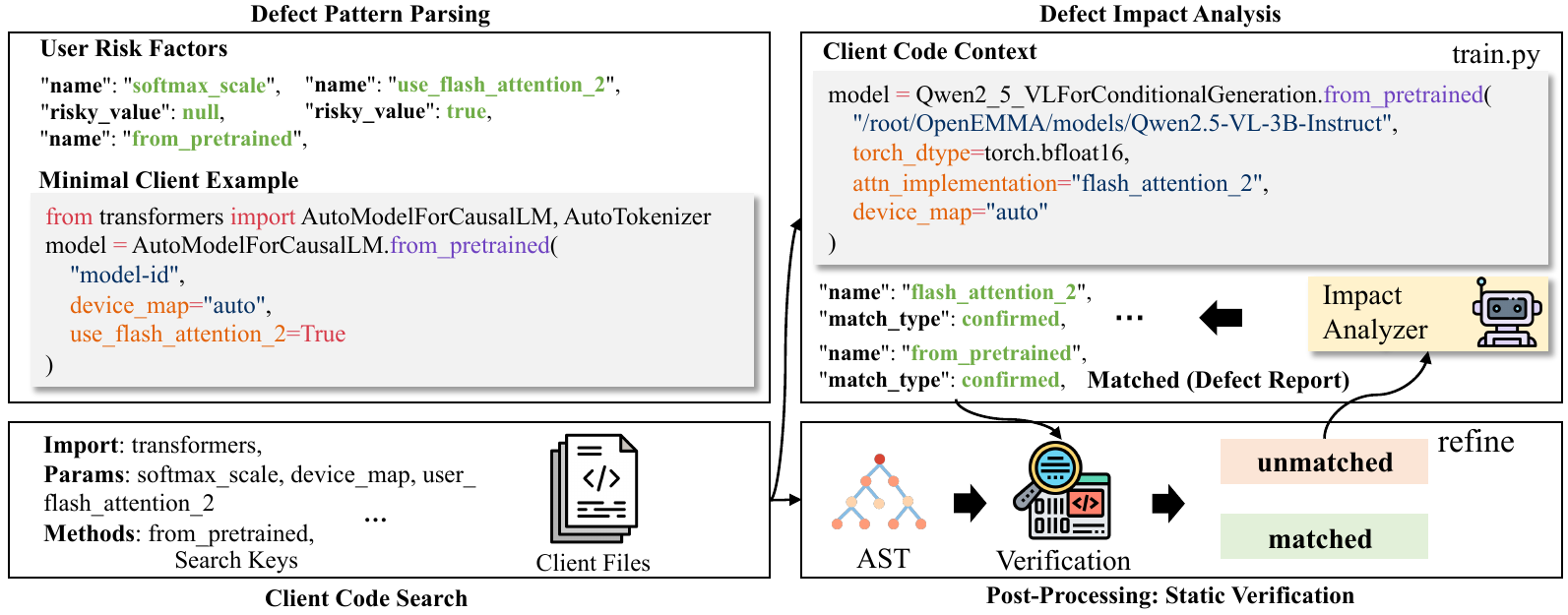}
\caption{Post-processing and verification: combining impact reports with AST analysis to validate methods and parameters.}
\label{fig:verify}
\vspace{-0.4cm}
\end{figure*}

\subsubsection{Version-Aware File Matching.} 
The first prerequisite for reliable impact analysis is verifying whether the client depends on a version of the DL library that contains the defect.
In Python, this is non-trivial due to varied dependency declarations across pip requirements or manually installed snapshots.
When a version is explicitly pinned (e.g., \texttt{transformers==4.30.2}), \appname aligns the PR/commit-level file changes with the target version via conservative AST-based subtree matching.

However, many projects omit strict versioning (e.g., \texttt{transfor mers\textgreater=4.0}), making resolution uncertain.
In such cases, \appname conservatively assumes potential exposure if the defect existed prior to the client’s runtime or install date—reflecting how \texttt{pip} typically resolves to the latest version.
This prevents false negatives in real deployments, where latent defects may propagate silently through frequent updates.
By this strategy, \appname automatically determines whether the client is likely to include the defect and continues to the impact analysis only when risk cannot be ruled out.

\subsubsection{Contextual Searching via Syntax-Aware Extraction.}
Given the defect pattern extracted in Step~2, \appname traverses the entire client code and constructs ASTs for each Python file. 
It then searches for occurrences of risk factors, such as specific parameter names, method calls, or triggering constants defined in the pattern.

Instead of analyzing the entire file, \appname extracts the minimal AST subtree that encloses each matched element (e.g., a model constructor or loop-level invocation). 
This localized context is then converted back to source code and passed to the Impact Analyzer Agent.
This syntax-aware extraction serves two purposes: (1) it preserves semantic locality by capturing only the immediate usage scenario, and (2) it reduces prompt size and noise, enabling focused reasoning. 
All extracted code regions are aggregated into a candidate context list for downstream defect impact analysis.

\noindent\textbf{Assessing Trigger Conditions.}
For each matched client context, the Impact Analyzer Agent determines whether the conditions required to activate the defect are satisfied. 
Figure~\ref{fig:analysis_agent} illustrates the prompt used for this analysis.
This includes checking the presence and values of risk-bearing parameters (e.g., \texttt{attn\_implementation ="flash\_attention\_2"}) and verifying that corresponding configurations are active in scope.

To handle variation in code style and naming, we incorporate domain-adapted prompting strategies that normalize synonyms (e.g., \texttt{enable\_flash\_attn\_v2} vs \texttt{use\_flash\_attention2}) and allow flexible argument ordering. 
If explicit conditions (e.g., use of GPU) are not contradicted in code, the agent conservatively assumes they hold.
This reasoning enables precise filtering: contexts that satisfy all trigger conditions proceed to reporting, while ambiguous or missing evidence leads to conservative rejection or re-analysis.

\noindent \textbf{Progressive Context Augmentation.}
Not all matched client contexts are sufficient for reliable impact assessment.
Similar to earlier agents, we adopt an adaptive multi-round inference loop coordinated by the Orchestrator Agent.
Each round starts with the minimal AST subtree containing the matched parameter or call.
If the agent’s reasoning yields incomplete or uncertain results—due to ambiguous control flow, shadowed variables, or overridden parameters—the Orchestrator expands the context upward along the AST, from the current node to enclosing blocks, functions, or classes.
All intermediate outputs are retained in memory for iterative refinement.
This process continues until either a confident conclusion is reached or no broader context remains.
It ensures both token efficiency and semantic completeness across diverse code structures.

\subsubsection{Post-processing and Verification.}
To mitigate the hallucinations commonly observed in LLM-generated predictions, we perform static validation by extracting predicted parameters and method calls (e.g., \texttt{softmax\_scale=None}) from the LLM output and verifying their presence in the client code. 
These predicted indicators are existence-checked against the client's source code using ASTs constructed via Python’s built-in \texttt{ast} module.
As illustrated in Figure~\ref{fig:verify}, the \appname checks whether each referenced element in the report—such as \texttt{use\_flash\_attention\_2} or specific API calls—can be confirmed by traversing the AST of the matched client context.
If any required evidence is missing, the orchestrator triggers a corrective re-analysis.
If the evidence remains insufficient after all retries, \appname conservatively marks the client as unaffected to avoid over-reporting.
This conservative strategy prioritizes precision by reducing false positives, ensuring that downstream impact is reported only when strongly supported by code evidence.

\subsubsection{Impact Report Generation.}
The final output of this step is a structured \textit{Impact Report}, which includes: (1) a binary \texttt{is\_affected} flag, (2) the set of matched parameters and method calls contributing to the judgment, and (3) a justification summary with both reasoning trace and static evidence. 
These reports are not only valuable for informing users of downstream risks, but also serve as input to potential mitigation strategies, such as patch recommendation or warning generation during CI workflows.

%% file: evaluation.tex
\section{Evaluation}
\subsection{Experimental Settings}
\noindent \textbf{Target Libraries.}
We evaluate \appname on two representative DL libraries: \textit{Transformers}~\cite{transformers} and \textit{Megatron}~\cite{megatron}.
Transformers is a widely used, community-maintained framework for general-purpose NLP and multimodal modeling, with over 20,000 pull requests and frequent feature updates.
\textit{Megatron}, by contrast, is an internally maintained \textit{NVIDIA}~\cite{nvidia} library optimized for large-scale distributed training; it has limited PR activity but continuously updated commits on its \texttt{main} branch.
These two libraries represent complementary scenarios: community-driven open development and enterprise-scale infrastructure evolution.
While \appname is generalizable to other scientific computing libraries, we focus on DL frameworks because their defects are often configuration-sensitive, tensor-dependent, and hardware-specific (e.g., GPU precision modes), making them representative targets for evaluating defect and impact analysis.

\noindent \textbf{PR/Commit Sampling and Annotation.}
For \textit{Transformers}, we collect all PRs merged during the Q2 2025 (April–June), obtaining 718 candidates.
We then perform a two-step procedure:
(1) \textit{Filtering}: PRs clearly unrelated to defect fixing (e.g., documentation, testing) are removed based on title and diff.
(2) \textit{Manual annotation}: the remaining 157 PRs are independently reviewed by two DL-domain experts, who decide whether the change is a \emph{real defect fix} or a benign change. 
Disagreements are later reconciled by discussion.
For \textit{Megatron}, since it does not actively maintain public PRs, we collect 535 recent commits from the \texttt{main} branch, exclude 465 non-defect cases, and retain 70 candidates.
These are annotated by the same two experts using identical criteria to ensure consistency across PR- and commit-based workflows.

\noindent \textbf{Baselines and Metrics.} 
We compare \appname against three representative baselines: FlatLLM (Base), FlatLLM (Reasoning), and PyCG~\cite{salis2021pycg}.
FlatLLM (Base) serves as a zero-shot summarization baseline~\cite{liu2019automatic,xiao2024generative}, reflecting the common practice of using a single general-purpose LLM to interpret PRs or commits without orchestration.
Although these models are primarily designed for commit or pull request summarization, they provide a realistic and conservative comparison for uncoordinated LLM reasoning over code changes.
We also compare \appname with FlatLLM (Reasoning), a self-reasoning baseline that uses the \texttt{DeepSeek-R1}~\cite{deepseekr1}. 
Unlike FlatLLM (Base), FlatLLM (Reasoning) is capable of reasoning internally and handling both defect pattern generation and impact analysis within the same model.
Finally, PyCG provides a non-LLM static call-graph baseline.
Unlike prior work on commit summarization using ROUGE~\cite{liu2019automatic,rouge}, our task produces structured representations (defect patterns) intended for downstream reasoning.
We therefore adopt expert-driven field-level scoring for summary evaluation and binary classification metrics (precision, recall, F1) for defect impact analysis.

\noindent \textbf{LLM Configuration.}
All agents in \appname use \texttt{DeepSeek-V3}~\cite{deepseek} as the backbone model.
We select \texttt{DeepSeek-V3} for its ability to efficiently process heterogeneous inputs (e.g., PR discussions, code diffs, and AST fragments) while maintaining stable reasoning quality at a manageable token cost.
To ensure reproducibility and avoid mixed-model effects, all agents share the same base model under deterministic decoding settings (temperature = 0, top-p = 1.0).

\noindent \textbf{Client Collection.}
\label{sec:clientcollection}
For PR analysis, we start from the 122 true-positive defect patterns and collect open-source downstream clients on GitHub that depend on \textit{Transformers}.
For each defect pattern, we extract key identifiers (e.g., model names, parameters, function calls such as \texttt{from\_pretrained()} and \texttt{attn\_implementation}) as query anchors to locate potential clients.
We retain only clients that can be successfully executed, pass integrity checks (e.g., dependency installation, model loading), and are not trivial forks or toy examples.
The resulting clients cover diverse downstream usage scenarios, including NLP fine-tuning pipelines, multimodal training frameworks, and inference services, representative of the broader \textit{Transformers} ecosystem.
For commit-level evaluation, we use \texttt{MindSpeed}~\cite{mindspeed}, a production-grade open-source project built on \textit{Megatron} for large-scale distributed training and deployment.
Its industrial workflow and extensive reuse of \textit{Megatron} modules make it an ideal environment for validating downstream impacts of commit-level fixes.

\subsection{RQ1: Can \appname effectively generate defect patterns?}
\noindent\textbf{Methodology.}
A key goal of \appname is to distinguish real bug-fix commits from benign changes and to extract precise, structured defect patterns from natural language and code changes.
To assess this capability, we evaluate \appname on 157 recent PRs from the \texttt{Transformers}. 
These PRs include a wide variety of silent defects—bugs that do not crash the program but can cause incorrect behavior or inefficiency under certain conditions.

\noindent\textbf{Results and Analysis.}  
As shown in Table~\ref{tab:rq1-bigtable}, \appname achieves high performance in defect identification, with a precision of 90\%, recall of 99\%, and F1-score of 95\%.
The specificity (i.e., true negative rate) is 62\%, indicating that \appname can also effectively distinguish non-defect changes such as documentation updates or test refactoring.
Compared to FlatLLM (base), which uses a single LLM summarization prompt without agentic coordination, \appname exhibits superior accuracy across all metrics, demonstrating the advantage of its structured reasoning and patch-aware analysis.
\appname correctly flags non-defect PRs such as \#38212 (fixing logging links) and \#38260 (adding decorators and docstrings) by recognizing their lack of semantic impact on runtime behavior. 
In contrast, the FlatLLM(base) misclassifies these cases due to the presence of keywords like \textit{fix} and superficial lexical changes.

\begin{table}[ht]
\centering
\small
\setlength{\tabcolsep}{3pt}
\caption{Defect Identification Performance Comparison.}
\label{tab:rq1-bigtable}
\begin{tabular}{lcccccccc}
\toprule
\textbf{Tool} & \textbf{TP} & \textbf{FP} & \textbf{FN} & \textbf{TN} & \textbf{Prec.} & \textbf{Rec.} & \textbf{F1} & \textbf{Spec.} \\
\midrule
\appname & 122 & 13 & 1 & 21 & \textbf{90\%} & \textbf{99\%} & \textbf{95\%} & \textbf{62\%} \\
FlatLLM (Base) & 112 & 27 & 10 & 7 & 81\% & 92\% & 86\% & 21\% \\
FlatLLM (Reasoning) & 116 & 19 & 6 & 15 & 86\% & 95\% & 90\% & 44\% \\
\bottomrule
\end{tabular}
\vspace{-0.3cm}
\end{table}
\vspace{0.2em}


While FlatLLM(Reasoning) improves recall over FlatLLM(Base) by generating intermediate reasoning chains, it often produces redundant or speculative interpretations when the patch semantics are ambiguous.
For example, in PR \#37905 fixing unintended quantization of tied embeddings, FlatLLM(Reasoning) misclassified the change as a performance optimization due to keyword overlap.
In contrast, \appname’s multi-agent coordination enables explicit semantic verification, correctly linking the quantization configuration to the tied-weight condition and identifying it as a real defect.

\noindent\textbf{Error Analysis.}
We manually inspect all false positives.
We find that 13 PRs are flagged as defects despite having no effect on core functionality.
These cases primarily involve bug fixes in test components.
For example, PR~\#38471 adjusts \texttt{FlexAttn} test logic to handle edge-case tensors in \texttt{Zamba2} and \texttt{Deepseek3} models, fixing a bug where \texttt{Gemma3IntegrationTest} was never executed.
While technically correcting a bug in test behavior, such patches do not influence production usage.
\appname currently lacks the ability to distinguish between test scaffolding and runtime code.
Incorporating learned classifiers or test directory heuristics could further reduce false positives in future iterations.

\begin{table}[ht]
\centering
\small
\setlength{\tabcolsep}{3pt}
\caption{Field-wise accuracy and inter-rater agreement of defect patterns (N = 157).}
\label{tab:field-score}
\begin{tabular}{lccccc}
\toprule
\textbf{Field} & \textbf{Miss (0)} & \textbf{Part. (1)} & \textbf{Acc. (2)} & \textbf{Avg.} & \textbf{Cohen’s $\kappa$} \\
\midrule
Bug Background & 31 (10\%) & 61 (19\%) & 222 (71\%) & 1.61 & 0.71 \\
Impact Scope & 11 (4\%) & 38 (12\%) & 265 (84\%) & 1.81 & 0.74 \\
Trigger Conditions & 34 (11\%) & 122 (39\%) & 158 (50\%) & 1.39 & 0.71 \\
\bottomrule
\end{tabular}
\vspace{-0.35cm}
\end{table}


\noindent\textbf{Structured Field Evaluation.}
Beyond binary classification, we further assess the quality of structured defect reports generated by \appname.
Each report includes three key fields—\texttt{bug\_background}, \texttt{impact\_scope}, and \texttt{trigger\_conditions}—that are essential for downstream impact analysis.
Two domain experts with over five years of DL development experience independently rated each field on a 3-point scale (0 = missing, 1 = partially correct, 2 = fully accurate).
The 3-point design balances granularity and reliability: binary labels overlook partially correct cases, whereas finer scales reduce consistency.
Each expert annotated all samples independently to prevent mutual influence.
Cohen’s Kappa averaged 0.72 (range 0.71–0.74), indicating substantial inter-rater agreement.
Disagreements (9\%) were discussed and resolved through reference to PR diffs and commit metadata until consensus was reached.
As shown in Table~\ref{tab:field-score}, \appname achieves high field-level accuracy:
over 70\% of the \texttt{bug\_background} and 84\% of the \texttt{impact\_scope} entries are rated fully accurate.
The \texttt{trigger\_conditions} field is comparatively harder, with partial scores in cases involving multiple parameters or environment dependencies.

\noindent\textbf{Defect Type Distribution.}
Modeling trigger conditions proves to be the most challenging.
In many PRs, preconditions are implicit—distributed across config defaults, internal logic, and usage assumptions.
For example, PR~\#37575’s behavior depends on both a device setting (NPU) and an unset scale parameter.
Capturing such cross-cutting logic requires synthesizing patch semantics, device context, and optional configurations.
Future directions include integrating execution traces or dependency graphs to further enhance condition extraction.

\begin{table*}
\centering
\caption{Defect Taxonomy and Examples in Transformers PRs.}
\label{tab:defect-categories}
\begin{tabular}{clcll}
\toprule
\textbf{Type} & \textbf{Name} & \textbf{Count} & \textbf{Description} & \textbf{Example PRs} \\
\midrule
A & Configuration Misuse & 16 & Incorrect parameter settings, misuse of optional arguments & \#38223, \#38288 \\
B & Missing Preconditions & 13 & Lack of boundary checks or unsafe fallback logic & \#38264, \#38945 \\
C & Semantic Logic Errors & 56 & Flawed control flow, tensor misalignment, wrong fallback logic & \#37704, \#38290 \\
D & Compatibility Issues & 23 & Incompatibilities due to API changes or module relocations & \#38841, \#39083 \\
E & Concurrency Bugs & 4 & Synchronization issues or data races across devices/threads & \#38557, \#38876 \\
F & Performance/Resource Bugs & 6 & Memory inefficiencies, unbounded CPU usage & \#38901, \#39043 \\
G & Crash or Fatal Errors & 4 & Crashes due to runtime exceptions or unsupported behavior & \#37690, \#38668 \\
\bottomrule
\end{tabular}
\vspace{-0.3cm}
\end{table*}


To assess whether \appname generalizes beyond narrow bug categories, we categorize all 122 true positive PRs into seven types (Table~\ref{tab:defect-categories}).
\appname achieves consistent performance across all categories, reflecting its robustness in modeling diverse defect semantics.
Notably, only four PRs belong to Type~G (crash), while over 97\% fall into non-crash categories.
These silent defects are difficult to detect via traditional testing, often requiring semantic patch interpretation and usage context reasoning.
The dominance of Type~C and D defects underscores the need for tools like \appname that move beyond crash detection and proactively reason about latent faults in modern DL systems.


\subsection{RQ2: How effective is \appname in analyzing whether a client is affected by a defect?}
\noindent\textbf{Methodology.}
After identifying bug-fix PRs and constructing structured defect patterns, a key challenge is to determine whether a downstream client program is genuinely \textit{affected}—i.e., whether it satisfies the trigger conditions of the defect and may exhibit incorrect behavior at runtime.
We evaluate \appname on 122 real-world \texttt{(defect, client)} pairs collected from the benchmark described in Sec.~\ref{sec:clientcollection}, and compare it against the three baselines.

\noindent\textbf{Results and Comparison.}
As shown in Table~\ref{tab:rq2-impact}, \appname achieves the best overall balance between precision and recall (\textbf{F1 = 85\%}), substantially outperforming all baselines.
Compared with FlatLLM(Base), which tends to over-report due to superficial lexical or contextual cues, \appname improves precision by 17 points (80\% vs.\ 63\%) and specificity by 28 points (72\% vs.\ 44\%).
FlatLLM(Reasoning) performs moderately better than Base (F1 = 74\% vs.\ 68\%) owing to its internal reasoning ability, but still suffers from false positives and missed configuration-dependent cases due to the lack of structural verification.
PyCG achieves the lowest overall performance (F1 = 47\%, Prec.\ = 44\%, Rec.\ = 50\%), as its static call-graph analysis misses dynamic dependencies and merges high-level APIs, causing many false positives and low recall.

\begin{table}[ht]
\centering
\small
\setlength{\tabcolsep}{3pt}
\caption{Impact Analysis Performance (N=122). The client set corresponds to TP defect patterns identified in RQ1.}
\label{tab:rq2-impact}
\begin{tabular}{lcccccccc}
\toprule
\textbf{Tool} & \textbf{TP} & \textbf{FP} & \textbf{FN} & \textbf{TN} & \textbf{Prec.} & \textbf{Rec.} & \textbf{F1} & \textbf{Spec.} \\
\midrule
\textbf{\appname}          & \textbf{61} & \textbf{15} & \textbf{7}  & \textbf{39} & \textbf{80\%} & \textbf{90\%} & \textbf{85\%} & \textbf{72\%} \\
FlatLLM (Base)              & 50          & 30          & 18          & 24          & 63\%         & 74\%          & 68\%          & 44\%          \\
FlatLLM (Reasoning)         & 54          & 25          & 14          & 29         & 68\%          & 79\%          & 74\%          & 54\%          \\
PyCG                       & 34          & 44          & 34          & 10          & 44\%          & 50\%          & 47\%          & 19\%         \\
\bottomrule
\end{tabular}
\vspace{-0.3cm}
\end{table}

\noindent\textbf{Qualitative Analysis.}
We find that \appname’s advantage stems from its ability to (1) \textit{match actual usage contexts in client code} and (2) \textit{verify their relevance through static AST checks}.
For example, in PR~\#38238, where the fix targets the configuration \texttt{decoder\_bbox \_embed\_share=False}, \appname correctly infers that a client using \texttt{GroundingDinoForObjectDetection} is \textbf{not affected} because the trigger condition is unmet—whereas both LLM baselines misclassify it as positive.
Similarly, for PR~\#38285, where the bug occurs inside a conditional macro, FlatLLM(Reasoning) fails to detect the impact due to implicit control flow, while \appname succeeds by structurally matching the AST predicates.


To further verify correctness, we construct test scripts for 50 randomly sampled TP cases and confirm that each exhibits the expected faulty behavior when executed under vulnerable library versions, such as memory bloat, incorrect attention scaling, or silent numerical drift.
These results show that \appname can go beyond surface-level call matching to semantically reason about trigger satisfaction in real-world usage. 
Its structured impact reports and context-aware analysis lead to both higher accuracy and more actionable diagnostics compared to prompt-only LLM summarization.

\subsection{RQ3: How does each component contribute to defect modeling and impact analysis?}
\subsubsection{Defect Pattern Generation Module Ablation} 
We perform an ablation study to evaluate the contribution of two key design dimensions in our approach: the use of \textit{coordinated framework} and the \textit{adaptive context expansion} mechanism for PRs.
Our goal is to quantify how each design choice affects the accuracy of defect pattern extraction.
To this end, we construct the dataset from Transformers PRs annotated in RQ1, and compare the following variants:
- \textbf{\appname (Full)}: Uses both the coordinated framework and adaptive context expansion.
- \textbf{w/o Adaptive Context}: Disables context expansion; each agent processes a fixed-size content window.
- \textbf{w/o Coordination}: Replaces multi-agent coordination  with a single LLM prompt that ingests all content (PR title, body, patch) at once, with no agent collaboration or role differentiation.

\begin{table}
\centering
\small
\setlength{\tabcolsep}{3pt}
\caption{Ablation Study on Defect Pattern Extraction (N=157).}
\label{tab:rq3-ablation}
\begin{tabular}{lccccccccc}
\toprule
\textbf{System Variant} & \textbf{TP} & \textbf{FP} & \textbf{FN} & \textbf{TN} & \textbf{Prec.} & \textbf{Rec.} & \textbf{F1} & \textbf{Spec.} \\
\midrule
\appname (Full)        &  122 & 13 &  1 & 21 & \textbf{90\%} & \textbf{99\%} & \textbf{95\%} & \textbf{62\%}  \\
w/o Coordination         &  78 & 27 & 45 &  7 & 74\%          & 63\%          & 68\%          & 21\%          \\
w/o Adaptive Context     & 118 & 17 &  5 & 17 & 87\%          & 96\%          & 91\%          & 50\%          \\
\bottomrule
\end{tabular}
\vspace{-0.4cm}
\end{table}


Table~\ref{tab:rq3-ablation} reports precision, recall, and F1 scores for defect classification. 
The full \appname achieves the highest performance (F1 = 95\%), while removing context adaptation causes a moderate decline (F1 = 91\%), primarily due to missed trigger conditions dispersed across PR metadata.
The single-prompt variant exhibits the largest drop (F1 = 68\%), as it lacks structured decomposition and often fails to disambiguate configuration details or code-specific semantics.
We find that this variant frequently hallucinates parameter names or overlooks patch semantics when processing longer inputs.

These results highlight that: (1) \textbf{adaptive context expansion} is crucial for capturing scattered defect signals across multi-round PR discussions, and (2) \textbf{multi-agent specialization} offers better grounding and modular control compared to flat summarization.
Together, these components enable \appname to robustly extract structured, actionable defect patterns across diverse bug types.

\subsubsection{Impact Analysis Module Ablation}
We evaluate the contribution of two critical modules in our pipeline: (1) the \textit{verification layer} that checks structural consistency between predicted risk factors and client code, and (2) the \textit{domain mapping layer} that maps internal library changes to user-facing constructs.

To prevent false positives from LLM hallucinations, \appname performs AST-based validation to confirm that predicted risk parameters and method calls are actually present in the matched client context. 
This layer also re-invokes the Impact Analyzer with feedback when inconsistencies arise.
Library patches often involve internal APIs (e.g., \texttt{scale\_mask}, \texttt{prepare\_inputs}) that users do not call directly. 
\appname encodes a set of DL-specific heuristics to project such internal signals to observable constructs—e.g., translating a buffer allocation bug to \texttt{AutoModel.from\_pretrained()} in user code. 
This mapping is crucial for matching abstract trigger conditions with real-world usage.

We construct two system variants:
- \textbf{w/o Verification Layer:} Skips consistency checks. The model’s first prediction is accepted without AST validation or retry logic.
- \textbf{w/o Domain Mapping:} Disables mapping from internal changes to user-facing methods and parameters.
Table~\ref{tab:rq3-impact} shows the performance impact on the Transformers client dataset.
Removing the verification layer significantly increases false positives, reducing precision from 80\% to 66\%.
Manual inspection reveals typical failure cases where the model hallucinates parameters such as \texttt{flash\_mask\_type}, which do not exist in the actual client code.
Without domain mapping, recall drops (90\%→76\%) as the model fails to align internal logic with surface-level usage.
For instance, in PR \#38257 (import error handling), the risk method \texttt{\_get\_module()} is never directly called by users, but its failure propagates through \texttt{AutoModel.from\_pretrained()}. Without mapping, this link is missed.

\begin{table}
\centering
\small
\setlength{\tabcolsep}{3pt}
\caption{Ablation Study on Client Impact Analysis (N = 122).}
\label{tab:rq3-impact}
\begin{tabular}{lcccccccc}
\toprule
\textbf{System Variant} & \textbf{TP} & \textbf{FP} & \textbf{FN} & \textbf{TN} & \textbf{Prec.} & \textbf{Rec.} & \textbf{F1} & \textbf{Spec.} \\
\midrule
\textbf{\appname (Full)}      & \textbf{61} & \textbf{15} & \textbf{7}  & \textbf{39} & \textbf{80\%} & \textbf{90\%} & \textbf{85\%} & \textbf{72\%} \\
w/o Verification Layer        & 59          & 30          & 9           & 24          & 66\%          & 87\%          & 75\%          & 44\%          \\
w/o Domain Mapping            & 52          & 22          & 16          & 32          & 70\%          & 76\%          & 73\%          & 59\%          \\
\bottomrule
\end{tabular}
\vspace{-0.4cm}
\end{table}


These results highlight the importance of grounded reasoning and domain abstraction in client impact analysis. 
Naively prompting LLMs—even with defect-aware context—is insufficient unless predictions are anchored in concrete code structure and adapted to real-world usage patterns.

\subsection{RQ4: Can \appname Effectively Handle Defect and Impact Analysis from Commits?}
\noindent\textbf{Methodology.}
We evaluate \appname on \textit{Megatron}, a commit-driven project without PRs, to examine its generalizability beyond PR-based workflows (Table~\ref{tab:rq4-merged}).
Unlike PR-based analysis, this setting evaluates whether \appname can maintain accuracy when only minimal commit information is available.

\noindent\textbf{Results.}
For defect pattern generation, it achieves a precision of 96\% and a recall of 90\%, despite the absence of high-level PR context. 
The few false negatives mainly correspond to commits involving test or documentation fixes that contain partial but non-functional code edits. 
This result indicates that \appname’s architecture—combining AST-based filtering, patch reasoning, and contextual alignment—retains high accuracy even when contextual cues are minimal.

\begin{table}[ht]
\centering
\setlength{\tabcolsep}{3pt}
\caption{Defect and Impact Analysis from Commits.}
\label{tab:rq4-merged}
\begin{tabular}{llcccccccc}
\toprule
\textbf{Task} & \textbf{TP} & \textbf{FP} & \textbf{FN} & \textbf{TN} & \textbf{Prec.} & \textbf{Rec.} & \textbf{F1} & \textbf{Spec.} \\
\midrule
\textbf{Defect Pattern} & 54 & 2 & 6 & 8 & \textbf{96\%} & \textbf{90\%} & \textbf{93\%} & 80\% \\
\textbf{Impact Analysis} & 12 & 8 & 5 & 29 & 60\% & 71\% & 65\% & 78\% \\
\bottomrule
\end{tabular}
\vspace{-0.4cm}
\end{table}


\noindent\textbf{Impact Analysis.}
For impact analysis, the task becomes considerably more challenging.
Unlike PRs, commits lack discussion and rationale context.
As a result, while recall remains high (the system still identifies most real impacts), precision declines due to increased false associations from superficial diffs.
Despite this sparsity, \appname still achieves a recall of 71\% and a precision of 60\%, successfully identifying 12 developer-confirmed client impacts among 54 true defect patterns.

\noindent\textbf{Developer-Confirmed Impacts.}
In the downstream client repository \texttt{MindSpeed}~\cite{mindspeed}, the development team validated \appname’s predictions, confirming 12 affected commits with real runtime impacts.
Among these, four were categorized as \textbf{high-risk} defects that directly degraded model accuracy or stability (e.g., gradient scaling), while the remaining eight were \textbf{low-risk} fixes such as logging inconsistencies, checkpoint I/O errors, or boundary-type corrections.
Table~\ref{tab:rq4-commits} summarizes representative developer-confirmed cases, illustrating how \appname distinguishes critical defects from benign propagation in industrial DL systems.

To further illustrate \appname’s effectiveness, we highlight two developer-confirmed cases from \textit{Megatron}.
In commit \textit{3bdcbbb Fix DDP scaling factor with Context Parallel}~\cite{3bdcbbb}, \textit{Megatron} developers fixed a bug in gradient scaling logic by adding a missing flag:
\begin{lstlisting}[language=Python]
# Original:
target_gradient_scaling_factor = 1.0 / parallel_state.get_data_parallel_world_size()

# Fixed:
target_gradient_scaling_factor = 1.0 / parallel_state.get_data_parallel_world_size(
with_context_parallel=True)
\end{lstlisting}

\appname successfully identified this as a silent numerical defect and generated a defect pattern with trigger conditions based on the usage of context parallelism.
In the downstream client \texttt{MindSpeed}~\cite{mindspeed}, which integrates \textit{Megatron} modules for large-scale model pretraining and deployment, a corresponding usage case was automatically matched through static analysis and flagged as \texttt{is\_affected: true}.
The developers later confirmed the bug's relevance and acknowledged its active impact on \texttt{MindSpeed}'s performance.
Instead of performing a full \textit{Megatron} upgrade (which would introduce compatibility issues), the developers implemented a localized patch incorporating the upstream fix - demonstrating the practical value of our impact analysis.

\begin{table}
\centering
\small
\caption{Developer-confirmed affected commits in \textit{Megatron}.}
\label{tab:rq4-commits}
\begin{tabular}{lll}
\toprule
\textbf{Risk} & \textbf{Commit} & \textbf{Primary Impact} \\
\midrule
High & \texttt{06e9f28}~\cite{06e9f28} & LR scaling bug under FSDP+TP mode \\
High & \texttt{1674ce3}~\cite{1674ce3} & Vision weight inconsistency under TP4 \\
High & \texttt{3bdcbbb}~\cite{3bdcbbb} & Gradient scaling instability (context parallel) \\
High & \texttt{dc4c5e7}~\cite{dc4c5e7} & Pipeline gradient accumulation mismatch \\
\midrule
Low & \texttt{8e2de56}~\cite{8e2de56} & Loss logging accumulation error \\
Low & \texttt{9a45638}~\cite{9a45638} & Optimizer checkpoint save/load fix \\
Low & \texttt{a77a883}~\cite{a77a883} & PP layer uneven split bugfix \\
Low & \texttt{db07e3f}~\cite{db07e3f} & Rotary-pos cosine check correction \\
\bottomrule
\end{tabular}
\vspace{-0.5cm}
\end{table}

Another case refers to \textit{Megatron} Commit \texttt{1674ce3}~\cite{1674ce3}.
This commit addresses a training defect in \textit{Megatron} where vision encoder weights were not correctly marked as non-tensor-parallelizable, leading to incorrect gradient reductions in TP-enabled training.
\appname automatically identified this commit as a real defect and extracted its root cause and trigger conditions. 
It further confirmed that \texttt{MindSpeed} is affected under TP4 configurations.
Empirical validation revealed that the bug caused gradient inconsistencies for unsharded vision weights during training and severe visual artifacts (\textit{screen tearing}) during inference.
To mitigate the issue, the \texttt{MindSpeed} team applied deterministic training mode to ensure consistent updates and is actively evaluating a fix by adding a new function \texttt{\_allreduce\_duplicate\_grads()} and modifying \texttt{finalize\_model\_grads()}.
Together, these results demonstrate that \appname can effectively uncover and contextualize silent defects that propagate into large-scale production systems, underscoring its robustness in real-world DL infrastructure.

\subsection{RQ5: What is the cost of using \appname?}
\noindent\textbf{Methodology.}
We analyze the computational efficiency of \appname when processing defect and impact analysis tasks, covering 157 PRs and 70 commits.

\noindent\textbf{Results.}
As shown in Table~\ref{tab:cost}, the \textit{Defect Pattern (PR)} task consumes about 5K tokens and 0.5 minutes per case, mainly due to the larger PR context.
\textit{Defect Pattern (Commit)} is lighter (<2K tokens, <0.3 min), while \textit{Impact Analysis (Client)} costs roughly 3K tokens and 1 minute per case due to multi-source reasoning.
These results show that token consumption and runtime scale with task complexity—PRs require richer contextual reasoning, while commits are lighter; client-side analysis incurs higher overhead due to multi-source dependency tracing.
Across all evaluated tasks, the total token consumption reaches approximately 1.29 million tokens, corresponding to a cost of about \$0.5 under \texttt{DeepSeek-V3}' pricing.
This demonstrates that even under conservative assumptions, the entire evaluation can be reproduced at negligible cost (under \$1).

\begin{table}[ht]
\centering
\small

\setlength{\tabcolsep}{3pt}
\caption{Computation cost and runtime of \appname across different task types. Tokens include both input and output.}
\label{tab:cost}
\resizebox{\columnwidth}{!}{
\begin{tabular}{lcccc}
\toprule
\textbf{Task Type} & \textbf{\makecell{Avg. \\ Tokens}} & \textbf{\makecell{Total \\ Tokens}}  & \textbf{\makecell{Avg. \\ Time (min)}} & \textbf{\makecell{Total \\ Time (min)}} \\
\midrule
Defect Pattern (PR)     & 5K   & 785K   & 0.5  & 78.5 \\
Defect Pattern (Commit) & <2K  & 140K   & <0.3 & 21  \\
Impact Analysis (Client)& 3K   & 366K   & 1.0  & 122 \\
\midrule
\textbf{Total} & - & \textbf{1.29M}  & - & \textbf{221.5} \\
\bottomrule
\end{tabular}
}
\vspace{-0.4cm}
\end{table}

\noindent\textbf{Comparison with Baselines.}
Compared with baseline models, the cost of \appname remains competitive.
\textit{FlatLLM (Base)} consumes a similar number of tokens, since its input context (PR body and code diffs) is comparable.
\textit{FlatLLM (Reasoning)} processes the same input but incurs higher expenses due to internal reasoning chains and longer inference times.
In contrast, \appname introduces only minor coordination overhead while achieving faster overall execution, as each agent operates within a focused sub-context.
The overall computational cost (in terms of tokens, runtime, and monetary expense) remains within a reasonable range.
Considering its analytical depth and coordination capability, \appname achieves an effective balance between efficiency and accuracy, demonstrating strong practicality for real-world deployment.

%% file: discussion.tex
\section{Discussion}
\label{sec:discussion}

\subsection{Generalizability, Adoption, and Extension}
\textbf{Generalizability.}
While our evaluation focuses on two popular DL libraries—Transformers and \textit{Megatron}, the modular architecture of \appname is broadly applicable to other Python-based libraries that expose public APIs and have structured commits or PRs.
We observe that \appname performs consistently across a range of defect types (e.g., configuration misuse, semantic errors, performance issues), suggesting good generalization. Extending the system to non-Python libraries (e.g., TensorFlow C++) would require adapting AST parsing and static analysis routines.

\noindent \textbf{Developer Adoption.}
Case studies in \textit{Megatron} show that users may prefer to locally backport patches instead of upgrading the entire library version, due to compatibility or dependency constraints. 
In such scenarios, \appname’s ability to surface concrete trigger conditions and risk factors can assist developers in writing minimal fixes. 
This highlights its potential utility as a defect-awareness tool in long-lived or production-grade codebases.

\noindent\textbf{Extension Directions.}
One promising extension is to integrate dynamic tracing or execution logs to better infer runtime-specific defects (e.g., those triggered by hardware mismatch or environment settings).
Another direction is model-agnostic tuning, where different LLMs (e.g., GPT-4, DeepSeek, Claude) are compared in terms of consistency, robustness, and hallucination rates under the same agent roles.
We also plan to explore IDE integration to offer actionable defect awareness during development or CI workflows.

\subsection{Threats to Validity}
\textbf{Threats to Internal Validity.} Some of our annotations—such as defect presence or client impact—rely on expert judgment, though we observe substantial inter-annotator agreement.
Certain trigger conditions may also depend on runtime paths not visible in static code.
Although \appname performs consistency checks and retry rounds, occasional hallucinations from LLMs cannot be fully ruled out.
\textbf{Threats to External Validity.} Our evaluation is limited to two actively maintained DL libraries.
Libraries with less structured development workflows or sparse metadata may degrade the effectiveness of agent-based reasoning.
In addition, the current system is designed for Python-based ecosystems; applying it to other languages requires further adaptation.

%% file: related_work.tex
\section{Related Work}
\label{related_work}
\textbf{Commit and Pull Request Understanding}
Understanding and analyzing PRs/commits is a fundamental task in software engineering research, aiming to support tasks such as code review, bug triage, and change impact analysis~\cite{icsik2025enhancing,chen2022untangling,li2022utango,dias2015untangling,muylaert2018untangling,herbold2022fine,liu2019automatic,xiao2024generative,bai2025clg,meng2024keytitle,meijer2024ecosystem,venigalla2024exploratory,pan2022automated}.
Liu et al.\cite{liu2019automatic} proposed a sequence-to-sequence model with a pointer-generator mechanism to automatically generate pull request (PR) descriptions by leveraging commit messages and code comments. 
They further applied reinforcement learning to directly optimize ROUGE scores and mitigate the mismatch between training objectives and evaluation metrics.
Bai et al.~\cite{bai2025clg} introduced the CLG framework, which generates PR checklists automatically from contribution guidelines by combining multi-label classification and text generation techniques. 
Their method significantly outperformed baselines on summarization tasks and demonstrated practical value in open-source community management.

\textbf{Security and Maintenance Challenges in Third-Party Library Dependencies.}
Prior work has extensively studied the risks posed by third-party libraries, including vulnerability propagation, delayed patch adoption, and ecosystem-wide dependency management challenges~\cite{zimmermann2019small,duan2020towards, huang2022characterizing,decan2019empirical,alfadel2023empirical,liu2022demystifying,chen2023identifying,zhang2023compatible,williams2025research,zhao2023software,zhang2023mitigating,zhang2024does,zhang2024vulnerability,gao2025vulnerability,zhou2024magneto,chen2025diffploitfacilitatingcrossversionexploit,liu2025empirical}. 
Zeng et al.~\cite{zeng2024survey} present a comprehensive survey on the security of third-party dependencies, highlighting trends in detection, mitigation, and secure integration practices.
Wu et al.~\cite{wu2023understanding} perform a large-scale empirical study over 44,450 triplets in the Maven ecosystem, revealing that 73.3\% of downstream projects depending on vulnerable libraries remain practically unaffected—underscoring the importance of precise impact assessment.
Chen et al.~\cite{chen2024exploiting} propose \textsc{Vesta} a vulnerability-oriented test generation framework that uses exploit code migration to evaluate real-world vulnerability reachability, improving detection success by 53.4\% over baselines.
Unlike prior work, \appname targets subtle, non-crashing defects in DL libraries by bridging defect modeling and downstream impact analysis through extracting fine-grained trigger conditions from code changes and semantically matching them to real client usage.

%% file: conclusion.tex
\section{Conclusion}
\label{conclusion}

In this paper, we presented \appname, an agent coordination framework for automating defect impact analysis in DL libraries.
\appname tackles the key challenge of determining whether bug fixes in libraries like \textit{Transformers} and \textit{Megatron} affect downstream client programs.
By coordinating Miner, Diff Analyzer, and Impact Analyzer agents—\appname bridges the semantic gap between low-level library patches and high-level user-facing behavior, enabling context-aware impact assessment. 
Our evaluation on real-world PRs and commits demonstrates its effectiveness, achieving up to 95\% F1 in defect classification and 80\% precision in impact analysis.
In the future, we plan to extend its capabilities in several directions: (1) broadening support for additional DL libraries (e.g., vllm~\cite{vllm}) and defect types, (2) scaling the evaluation to larger datasets of client programs.
While the current workflow focuses on PRs and commits containing explicit fixes, future extensions may also consider report-only or reproduction-case scenarios.
We further plan to extend \appname to security vulnerability analysis, enabling it can interpret CVE reports and perform structured impact reasoning similar to defect cases.

%% file: main.bib
@misc{depradar,
  title={DepRadar},year={2026},
  howpublished={\url{https://github.com/testmigrator/DepRadar}},
}

@misc{transformers,
  title={Transformers},year={2025},
  howpublished={\url{https://github.com/huggingface/transformers}},
}

@misc{megatron,
  title={Megatron-LM},year={2025},
  howpublished={\url{https://github.com/NVIDIA/Megatron-LM}},
}

@misc{autogen,
  title={AutoGen},year={2025},
  howpublished={\url{https://microsoft.github.io/autogen/stable/index.html}},
}

@misc{nvidia,
  title={NVIDIA},year={2025},
  howpublished={\url{https://www.nvidia.com/en-sg/}},
}

@misc{mindspeed,
  title={MindSpeed},year={2025},
  howpublished={\url{https://gitee.com/ascend/MindSpeed}},
}

@misc{rouge,
  title={ROUGE},year={2025},
  howpublished={\url{https://en.wikipedia.org/wiki/ROUGE_(metric)}},
}

@misc{deepseek,
  title={DeepSeek},year={2025},
  howpublished={\url{https://www.deepseek.com/}},
}

@misc{deepseekr1,
  title={DeepSeekR1},year={2025},
  howpublished={\url{https://github.com/deepseek-ai/DeepSeek-R1}},
}

@misc{vllm,
  title={vLLM},year={2025},
  howpublished={\url{https://github.com/vllm-project/vllm}},
}

@misc{3bdcbbb,
  title={MegatronCommit3bdcbbb},year={2025},
  howpublished={\url{https://github.com/NVIDIA/Megatron-LM/commit/3bdcbbbe5d2a455a75e28969be7250cd4bd27bae}},
}

@misc{1674ce3,
  title={MegatronCommit1674ce3},year={2025},
  howpublished={\url{https://github.com/NVIDIA/Megatron-LM/commit/1674ce30b82a90a729cc7bf2f93b41da47c0983e}},
}

@misc{dc4c5e7,
  title={MegatronCommitdc4c5e7},year={2025},
  howpublished={\url{https://github.com/NVIDIA/Megatron-LM/commit/dc4c5e71c734e2e72a02d6f7b9851ba9b4c80706}},
}

@misc{06e9f28,
  title={MegatronCommit06e9f28},year={2025},
  howpublished={\url{https://github.com/NVIDIA/Megatron-LM/commit/06e9f28b30e55a1c857784696626f9a703450d97}},
}

@misc{9a45638,
  title={MegatronCommit9a45638},year={2025},
  howpublished={\url{https://github.com/NVIDIA/Megatron-LM/commit/9a45638758de6bb6d1a2951185a2d5999e1dad92}},
}

@misc{8e2de56,
  title={MegatronCommit8e2de56},year={2025},
  howpublished={\url{https://github.com/NVIDIA/Megatron-LM/commit/8e2de5622b8df222e7c71e04c85670e8f3cf457c}},
}

@misc{db07e3f,
  title={MegatronCommitdb07e3f},year={2025},
  howpublished={\url{https://github.com/NVIDIA/Megatron-LM/commit/db07e3fe81e0fe7e0f12bab374830823fd2333b5}},
}

@misc{a77a883,
  title={MegatronCommita77a883},year={2025},
  howpublished={\url{https://github.com/NVIDIA/Megatron-LM/commit/a77a883e248e68df1912df4ef2cf05b712947fce}},
}

@misc{pr,
  title={PullRequests},year={2025},
  howpublished={\url{https://docs.github.com/en/pull-requests}},
}

@misc{commit,
  title={GitCommit},year={2025},
  howpublished={\url{https://github.com/git-guides/git-commit}},
}

@inproceedings{icsik2025enhancing,
  title={Enhancing Pull Request Reviews: Leveraging Large Language Models to Detect Inconsistencies Between Issues and Pull Requests},
  author={I{\c{s}}{\i}k, Ali Tunahan and {\c{C}}a{\u{g}}lar, Hatice K{\"u}bra and T{\"u}z{\"u}n, Eray},
  booktitle={2025 IEEE/ACM Second International Conference on AI Foundation Models and Software Engineering (Forge)},
  pages={168--178},
  year={2025},
  organization={IEEE}
}

@inproceedings{chen2022untangling,
  title={Untangling composite commits by attributed graph clustering},
  author={Chen, Siyu and Xu, Shengbin and Yao, Yuan and Xu, Feng},
  booktitle={Proceedings of the 13th Asia-Pacific Symposium on Internetware},
  pages={117--126},
  year={2022}
}

@inproceedings{li2022utango,
  title={UTANGO: untangling commits with context-aware, graph-based, code change clustering learning model},
  author={Li, Yi and Wang, Shaohua and Nguyen, Tien N},
  booktitle={Proceedings of the 30th ACM Joint European Software Engineering Conference and Symposium on the Foundations of Software Engineering},
  pages={221--232},
  year={2022}
}

@inproceedings{dias2015untangling,
  title={Untangling fine-grained code changes},
  author={Dias, Mart{\'\i}n and Bacchelli, Alberto and Gousios, Georgios and Cassou, Damien and Ducasse, St{\'e}phane},
  booktitle={2015 IEEE 22nd International Conference on Software Analysis, Evolution, and Reengineering (SANER)},
  pages={341--350},
  year={2015},
  organization={IEEE}
}

@inproceedings{muylaert2018untangling,
  title={Untangling composite commits using program slicing},
  author={Muylaert, Ward and De Roover, Coen},
  booktitle={2018 IEEE 18th International Working Conference on Source Code Analysis and Manipulation (SCAM)},
  pages={193--202},
  year={2018},
  organization={IEEE}
}

@article{herbold2022fine,
  title={A fine-grained data set and analysis of tangling in bug fixing commits},
  author={Herbold, Steffen and Trautsch, Alexander and Ledel, Benjamin and Aghamohammadi, Alireza and Ghaleb, Taher A and Chahal, Kuljit Kaur and Bossenmaier, Tim and Nagaria, Bhaveet and Makedonski, Philip and Ahmadabadi, Matin Nili and others},
  journal={Empirical Software Engineering},
  volume={27},
  number={6},
  pages={125},
  year={2022},
  publisher={Springer}
}

@inproceedings{liu2019automatic,
  title={Automatic generation of pull request descriptions},
  author={Liu, Zhongxin and Xia, Xin and Treude, Christoph and Lo, David and Li, Shanping},
  booktitle={2019 34th IEEE/ACM International Conference on Automated Software Engineering (ASE)},
  pages={176--188},
  year={2019},
  organization={IEEE}
}

@article{xiao2024generative,
  title={Generative AI for pull request descriptions: Adoption, impact, and developer interventions},
  author={Xiao, Tao and Hata, Hideaki and Treude, Christoph and Matsumoto, Kenichi},
  journal={Proceedings of the ACM on Software Engineering},
  volume={1},
  number={FSE},
  pages={1043--1065},
  year={2024},
  publisher={ACM New York, NY, USA}
}

@article{venigalla2024exploratory,
  title={An exploratory study of software artifacts on GitHub from the lens of documentation},
  author={Venigalla, Akhila Sri Manasa and Chimalakonda, Sridhar},
  journal={Information and Software Technology},
  volume={169},
  pages={107425},
  year={2024},
  publisher={Elsevier}
}

@article{meijer2024ecosystem,
  title={Ecosystem-wide influences on pull request decisions: insights from NPM},
  author={Meijer, Willem and Riveni, Mirela and Rastogi, Ayushi},
  journal={arXiv preprint arXiv:2410.14695},
  year={2024}
}

@article{bai2025clg,
  title={CLG: Automated checklist generation for improved pull request quality},
  author={Bai, Shuotong and Meng, Chenkun and Li, Guodong and Liu, Huaxiao and Liu, Lei},
  journal={Expert Systems with Applications},
  volume={277},
  pages={127178},
  year={2025},
  publisher={Elsevier}
}

@article{meng2024keytitle,
  title={KeyTitle: towards better bug report title generation by keywords planning},
  author={Meng, Qianshuang and Zou, Weiqin and Cai, Biyu and Zhang, Jingxuan},
  journal={Software Quality Journal},
  volume={32},
  number={4},
  pages={1655--1682},
  year={2024},
  publisher={Springer}
}

@misc{chen2025diffploitfacilitatingcrossversionexploit,
      title={Diffploit: Facilitating Cross-Version Exploit Migration for Open Source Library Vulnerabilities}, 
      author={Zirui Chen and Zhipeng Xue and Jiayuan Zhou and Xing Hu and Xin Xia and Xiaohu Yang},
      year={2025},
      eprint={2511.12950},
      archivePrefix={arXiv},
      primaryClass={cs.SE},
      url={https://arxiv.org/abs/2511.12950}, 
}

@article{zeng2024survey,
  title={A Survey of Third-Party Library Security Research in Application Software},
  author={Zeng, Jia and Han, Dan and Zhu, Yaling and Wang, Yangzhong and Weng, Fangchen},
  journal={arXiv preprint arXiv:2404.17955},
  year={2024}
}

@inproceedings{wu2023understanding,
  title={Understanding the Threats of Upstream Vulnerabilities to Downstream Projects in the Maven Ecosystem},
  author={Wu, Yulun and Yu, Zeliang and Wen, Ming and Li, Qiang and Zou, Deqing and Jin, Hai},
  booktitle={2023 IEEE/ACM 45th International Conference on Software Engineering (ICSE)},
  pages={1046--1058},
  year={2023},
  organization={IEEE}
}

@article{lyu2023chronos,
  title={Chronos: Time-aware zero-shot identification of libraries from vulnerability reports},
  author={Lyu, Yunbo and Le-Cong, Thanh and Kang, Hong Jin and Widyasari, Ratnadira and Zhao, Zhipeng and Le, Xuan-Bach D and Li, Ming and Lo, David},
  journal={arXiv preprint arXiv:2301.03944},
  year={2023}
}

@article{kula2018developers,
  title={Do developers update their library dependencies? An empirical study on the impact of security advisories on library migration},
  author={Kula, Raula Gaikovina and German, Daniel M and Ouni, Ali and Ishio, Takashi and Inoue, Katsuro},
  journal={Empirical Software Engineering},
  volume={23},
  pages={384--417},
  year={2018},
  publisher={Springer}
}

@article{samarinunderstanding,
  title={Understanding How Third-Party Libraries in Mobile Apps Affect Responses to Subject Access Requests},
  author={Samarin, Nikita and Wijesekera, Primal}
}

@article{chen2023toward,
  title={Toward understanding deep learning framework bugs},
  author={Chen, Junjie and Liang, Yihua and Shen, Qingchao and Jiang, Jiajun and Li, Shuochuan},
  journal={ACM Transactions on Software Engineering and Methodology},
  volume={32},
  number={6},
  pages={1--31},
  year={2023},
  publisher={ACM New York, NY}
}

@article{yu2025towards,
  title={Towards Understanding Bugs in Distributed Training and Inference Frameworks for Large Language Models},
  author={Yu, Xiao and Chen, Haoxuan and Niu, Feifei and Hu, Xing and Keung, Jacky Wai and Xia, Xin},
  journal={arXiv preprint arXiv:2506.10426},
  year={2025}
}

@inproceedings{chen2024exploiting,
  title={Exploiting library vulnerability via migration based automating test generation},
  author={Chen, Zirui and Hu, Xing and Xia, Xin and Gao, Yi and Xu, Tongtong and Lo, David and Yang, Xiaohu},
  booktitle={Proceedings of the IEEE/ACM 46th International Conference on Software Engineering},
  pages={1--12},
  year={2024}
}

@article{duan2020towards,
  title={Towards measuring supply chain attacks on package managers for interpreted languages},
  author={Duan, Ruian and Alrawi, Omar and Kasturi, Ranjita Pai and Elder, Ryan and Saltaformaggio, Brendan and Lee, Wenke},
  journal={arXiv preprint arXiv:2002.01139},
  year={2020}
}

@article{huang2022characterizing,
  title={Characterizing usages, updates and risks of third-party libraries in Java projects},
  author={Huang, Kaifeng and Chen, Bihuan and Xu, Congying and Wang, Ying and Shi, Bowen and Peng, Xin and Wu, Yijian and Liu, Yang},
  journal={Empirical Software Engineering},
  volume={27},
  number={4},
  pages={90},
  year={2022},
  publisher={Springer}
}

@article{decan2019empirical,
  title={An empirical comparison of dependency network evolution in seven software packaging ecosystems},
  author={Decan, Alexandre and Mens, Tom and Grosjean, Philippe},
  journal={Empirical Software Engineering},
  volume={24},
  pages={381--416},
  year={2019},
  publisher={Springer}
}

@inproceedings{zimmermann2019small,
  title={Small world with high risks: A study of security threats in the npm ecosystem},
  author={Zimmermann, Markus and Staicu, Cristian-Alexandru and Tenny, Cam and Pradel, Michael},
  booktitle={28th USENIX Security Symposium (USENIX Security 19)},
  pages={995--1010},
  year={2019}
}

@article{alfadel2023empirical,
  title={Empirical analysis of security vulnerabilities in python packages},
  author={Alfadel, Mahmoud and Costa, Diego Elias and Shihab, Emad},
  journal={Empirical Software Engineering},
  volume={28},
  number={3},
  pages={59},
  year={2023},
  publisher={Springer}
}

@inproceedings{liu2022demystifying,
  title={Demystifying the vulnerability propagation and its evolution via dependency trees in the npm ecosystem},
  author={Liu, Chengwei and Chen, Sen and Fan, Lingling and Chen, Bihuan and Liu, Yang and Peng, Xin},
  booktitle={Proceedings of the 44th International Conference on Software Engineering},
  pages={672--684},
  year={2022}
}

@misc{chen2023identifying,
      title={Identifying Vulnerable Third-Party Libraries from Textual Descriptions of Vulnerabilities and Libraries}, 
      author={Tianyu Chen and Lin Li and Bingjie Shan and Guangtai Liang and Ding Li and Qianxiang Wang and Tao Xie},
      year={2023},
      eprint={2307.08206},
      archivePrefix={arXiv},
      primaryClass={cs.CR}
}

@misc{zhang2023compatible,
      title={Compatible Remediation on Vulnerabilities from Third-Party Libraries for Java Projects}, 
      author={Lyuye Zhang and Chengwei Liu and Zhengzi Xu and Sen Chen and Lingling Fan and Lida Zhao and Jiahui Wu and Yang Liu},
      year={2023},
      eprint={2301.08434},
      archivePrefix={arXiv},
      primaryClass={cs.SE}
}

@article{williams2025research,
  title={Research directions in software supply chain security},
  author={Williams, Laurie and Benedetti, Giacomo and Hamer, Sivana and Paramitha, Ranindya and Rahman, Imranur and Tamanna, Mahzabin and Tystahl, Greg and Zahan, Nusrat and Morrison, Patrick and Acar, Yasemin and others},
  journal={ACM Transactions on Software Engineering and Methodology},
  volume={34},
  number={5},
  pages={1--38},
  year={2025},
  publisher={ACM New York, NY}
}

@inproceedings{zhao2023software,
  title={Software composition analysis for vulnerability detection: An empirical study on Java projects},
  author={Zhao, Lida and Chen, Sen and Xu, Zhengzi and Liu, Chengwei and Zhang, Lyuye and Wu, Jiahui and Sun, Jun and Liu, Yang},
  booktitle={Proceedings of the 31st ACM Joint European Software Engineering Conference and Symposium on the Foundations of Software Engineering},
  pages={960--972},
  year={2023}
}

@inproceedings{zhang2023mitigating,
  title={Mitigating persistence of open-source vulnerabilities in maven ecosystem},
  author={Zhang, Lyuye and Liu, Chengwei and Chen, Sen and Xu, Zhengzi and Fan, Lingling and Zhao, Lida and Zhang, Yiran and Liu, Yang},
  booktitle={2023 38th IEEE/ACM International Conference on Automated Software Engineering (ASE)},
  pages={191--203},
  year={2023},
  organization={IEEE}
}

@article{zhang2024does,
  title={Does the vulnerability threaten our projects? automated vulnerable api detection for third-party libraries},
  author={Zhang, Fangyuan and Fan, Lingling and Chen, Sen and Cai, Miaoying and Xu, Sihan and Zhao, Lida},
  journal={IEEE Transactions on Software Engineering},
  year={2024},
  publisher={IEEE}
}

@inproceedings{zhang2024vulnerability,
  title={Vulnerability root cause function locating for java vulnerabilities},
  author={Zhang, Lyuye},
  booktitle={Proceedings of the 2024 IEEE/ACM 46th International Conference on Software Engineering: Companion Proceedings},
  pages={444--446},
  year={2024}
}

@inproceedings{gao2025vulnerability,
  title={Vulnerability-Triggering Test Case Generation from Third-Party Libraries},
  author={Gao, Yi and Hu, Xing and Chen, Zirui and Xu, Tongtong and Yang, Xiaohu},
  booktitle={2025 IEEE/ACM Second International Conference on AI Foundation Models and Software Engineering (Forge)},
  pages={125--135},
  year={2025},
  organization={IEEE}
}

@inproceedings{zhou2024magneto,
  title={Magneto: A Step-Wise Approach to Exploit Vulnerabilities in Dependent Libraries via LLM-Empowered Directed Fuzzing},
  author={Zhou, Zhuotong and Yang, Yongzhuo and Wu, Susheng and Huang, Yiheng and Chen, Bihuan and Peng, Xin},
  booktitle={Proceedings of the 39th IEEE/ACM International Conference on Automated Software Engineering},
  pages={1633--1644},
  year={2024}
}

@inproceedings{pepe2024taxonomy,
  title={A taxonomy of self-admitted technical debt in deep learning systems},
  author={Pepe, Federica and Zampetti, Fiorella and Mastropaolo, Antonio and Bavota, Gabriele and Di Penta, Massimiliano},
  booktitle={2024 IEEE International Conference on Software Maintenance and Evolution (ICSME)},
  pages={388--399},
  year={2024},
  organization={IEEE}
}

@article{li2024knowbug,
  title={KnowBug: Enhancing Large language models with bug report knowledge for deep learning framework bug prediction},
  author={Li, Chenglong and Zheng, Zheng and Du, Xiaoting and Ma, Xiangyue and Wang, Zhengqi and Li, Xinheng},
  journal={Knowledge-Based Systems},
  volume={305},
  pages={112588},
  year={2024},
  publisher={Elsevier}
}

@article{jia2021symptoms,
  title={The symptoms, causes, and repairs of bugs inside a deep learning library},
  author={Jia, Li and Zhong, Hao and Wang, Xiaoyin and Huang, Linpeng and Lu, Xuansheng},
  journal={Journal of Systems and Software},
  volume={177},
  pages={110935},
  year={2021},
  publisher={Elsevier}
}

@inproceedings{otoom2019automated,
  title={Automated classification of software bug reports},
  author={Otoom, Ahmed Fawzi and Al-jdaeh, Sara and Hammad, Maen},
  booktitle={Proceedings of the 9th international conference on information communication and management},
  pages={17--21},
  year={2019}
}

@article{2024Understanding,
  title={Understanding the implementation issues when using deep learning frameworks},
  author={ Liu, C.  and  Cai, R.  and Zhou Y.Chen X.Hu H.Yan M.},
  journal={Information and software technology},
  volume={166},
  number={Feb.},
  pages={1.1-1.16},
  year={2024},
}

@article{2023Toward,
  title={Toward Understanding Deep Learning Framework Bugs},
  author={ Chen, Junjie  and  Liang, Yihua  and  Shen, Qingchao  and  Jiang, Jiajun  and  Li, Shuochuan },
  journal={ACM Transactions on Software Engineering and Methodology},
  volume={32},
  number={6},
  pages={31},
  year={2023},
}

@article{2024Bug,
  title={Bug characterization in machine learning-based systems},
  author={ Morovati, Mohammad Mehdi  and  Nikanjam, Amin  and  Jiang, Khomh Zhen Ming Jack },
  journal={Empirical software engineering},
  volume={29},
  number={1},
  pages={1.1-1.29},
  year={2024},
}

@inproceedings{10.5555/3767901.3767919,
author = {Jiang, Yuxuan and Zhou, Ziming and Xu, Boyu and Liu, Beijie and Xu, Runhui and Huang, Peng},
title = {Training with confidence: catching silent errors in deep learning training with automated proactive checks},
year = {2025},
isbn = {978-1-939133-47-2},
publisher = {USENIX Association},
address = {USA},
booktitle = {Proceedings of the 19th USENIX Conference on Operating Systems Design and Implementation},
articleno = {18},
numpages = {17},
location = {Boston, MA, USA},
series = {OSDI '25}
}

@inproceedings{salis2021pycg,
  title={Pycg: Practical call graph generation in python},
  author={Salis, Vitalis and Sotiropoulos, Thodoris and Louridas, Panos and Spinellis, Diomidis and Mitropoulos, Dimitris},
  booktitle={2021 IEEE/ACM 43rd International Conference on Software Engineering (ICSE)},
  pages={1646--1657},
  year={2021},
  organization={IEEE}
}

@inproceedings{pan2022automated,
  title={Automated unearthing of dangerous issue reports},
  author={Pan, Shengyi and Zhou, Jiayuan and Cogo, Filipe Roseiro and Xia, Xin and Bao, Lingfeng and Hu, Xing and Li, Shanping and Hassan, Ahmed E},
  booktitle={Proceedings of the 30th ACM Joint European Software Engineering Conference and Symposium on the Foundations of Software Engineering},
  pages={834--846},
  year={2022}
}

@article{liu2025empirical,
  title={An Empirical Study of Vulnerable Package Dependencies in LLM Repositories},
  author={Liu, Shuhan and Hu, Xing and Xia, Xin and Lo, David and Yang, Xiaohu},
  journal={arXiv preprint arXiv:2508.21417},
  year={2025}
}
